\documentclass{aa}
\usepackage{graphicx}
\usepackage{txfonts}
\usepackage{lipsum}
\usepackage{subcaption}
\usepackage{placeins}
\usepackage{hyperref}
\begin{document}
\let\linenumbers\relax
\let\nolinenumbers\relax
\title{Shape, regolith size and thickness, SMFe$^0$ content, and spectral type of Tianwen-2 target asteroid (469219) Kamo`oalewa}

\subtitle{}

\author{Pengfei Zhang\inst{1}\fnmsep\thanks{These authors contributed equally to this work.}   
          \and Guozheng Zhang\inst{2}\fnmsep\footnotemark[1]
          \and Yongxiong Zhang \inst{3}
          \and Marco Fenucci \inst {4,5}
          \and Pierre Vernazza \inst {6}
          \and Jin Zhao \inst{7}
          \and Yunbo Niu \inst{8}
          \and Xuejin Lu \inst{9}   
          \and Xing Wu\inst{10}   
          \and Honglei Lin\inst{11}
          \and Edward Cloutis\inst{12}
          \and Xiaoran Yan\inst{13}
          \and Xiaoping Lu\inst{14}
          \and Xiaobin Wang\inst{15}
          \and Xiaoping Zhang\inst{2}
          \and Yang Li\inst{1}\fnmsep\thanks{Corresponding author. liyang@mail.gyig.ac.cn}
        }

\institute{Center for Lunar and Planetary Sciences, Institute of Geochemistry, Chinese Academy of Sciences, Guiyang (Guizhou), China
             \and State Key Laboratory of Lunar and Planetary Sciences, Macau University of Science and Technology, Macau, China
             \and School of Engineering, Guangzhou College of Technology and Business, Guangzhou (Guangdong), China
             \and ESA ESRIN/PDO/NEO Coordination Centre, Largo Galileo Galilei, 1, Frascati (RM), Italy
             \and Deimos Italia, Via Giuseppe Verdi, 6, San Pietro Mosezzo (NO), Italy
             \and Aix Marseille Université, CNRS, CNES, Laboratoire d’Astrophysique de Marseille, Marseille, France
             \and Institute of Analysis and testing, Beijing Academy of Science and Technology, Beijing Center for Physical \& Chemical Analysis, Beijing, China
             \and Chongqing University State Key Laboratory of Power Transmission Equipment and System Security and New Technology, Institute of Systems Engineering, Chongqing, China
             \and Shandong Key Laboratory of Optical Astronomy and Solar-Terrestrial Environment, School of Space Science and Physics, Institute of Space Sciences, Shandong University, Weihai (Shandong), China
             \and State Key Laboratory of Space Weather, National Space Science Center, Chinese Academy of Sciences, Beijing, China
             \and Key Laboratory of the Earth and Planetary Physics, Institute of Geology and Geophysics, Chinese Academy of Sciences, Beijing, China
             \and Department of Geography, University of Winnipeg, Winnipeg, Canada
             \and Istituto di Fisica Applicata ``Nello Carrara” (IFAC-CNR), Sesto Fiorentino (FI), Italy
             \and School of Computer Science and Engineering, Faculty of Innovation Engineering, Macau University of Science and Technology, Macau, China
             \and Yunnan Observatories, Chinese Academy of Sciences, Kunming (Yunnan), China}   
             
\date{Received March 4, 2025 / Accepted XX XX, 2026}
\abstract
{China's Tianwen-2 spacecraft will return samples from the near-Earth asteroid (469219) Kamo`oalewa. We previously reported that Kamo`oalewa develops an LL-chondrite-compositional, highly space-weathered surface.}
{This study aims to estimate Kamo`oalewa's shape, regolith grain size and thickness, sub-micrometer iron (SMFe$^0$) content, and spectral type.}
{Using the lightcurve data and the Cellinoid model, we modeled Kamo`oalewa's shape, rotation period, and pole orientation. We then estimated its global distribution of regolith critical size using the balance method of gravity, cohesive force, and centrifugal force. Furthermore, in the temperature range of 253.15 to 473.15~K, we measured the thermal parameters of laser-irradiated LL chondrite powder that best matches Kamo`oalewa's spectrum, estimating Kamo`oalewa's thermal inertia and skin depth (lower limit of regolith thickness). Using the radiative transfer mixing model, we also estimated the content of SMFe$^0$ in Kamo`oalewa's regolith. Finally, using the MIT online spectral classification tool for the laser-irradiated LL chondrite powder, we obtained a virtual spectral type of Kamo`oalewa.}
{Our model gives a size of 68~m $\times$ 46~m $\times$ 39~m, a rotation period of 27.66~minutes, and a pole orientation of 134.7$^{\circ}$ longitude and $-$11.4$^{\circ}$ latitude for kamo`oalewa. Regolith grains with a size <2~cm can remain stable over 93.8\% of the global surface area of Kamo`oalewa. Laser-irradiated LL chondrite powder shows a low thermal inertia (95.5 to 135.1 J m$^{-2}$ K$^{-1}$ s$^{-1/2}$), corresponding to a thermal skin depth of 3 to 3.5~mm on Kamo`oalewa. An SMFe$^0$ content of 0.29 $\pm$ 0.05 wt.\% is required to match Kamo`oalewa's spectrum. The virtual spectral type of Kamo`oalewa is given as ``Sqw''.}
{}


\authorrunning{Pengfei Zhang et al.}
\titlerunning{Shape, regolith size and thickness, SMFe$^0$ content, and spectral type of Kamo`oalewa}

   \maketitle
\section{Introduction}\label{1}
Asteroids, as remnants of the planetary accretion process, record information on the early Solar System's environment and subsequent evolutionary history. For this reason, a large number of ground-based \citep{chapman1979reflectance,zellner1985eight,bell198852,bus2002phasea,carvano2010sdss, de2018primitive,binzel2019compositional,sanchez2024population} and space-based \citep{galluccio2023gaia} spectroscopic survey projects and and on-orbit detection mission (such as NEAR Shoemaker to 433 Eros \citep{prockter2002near}, Dawn to 1 Ceres and 4 Vesta \citep{russell2011dawn}, Lucy to Jupiter Trojan asteroids \citep{levison2021lucy}, and Psyche to metal-rich asteroid 16 Psyche \citep{lord2017psyche}), have been carried out. These have greatly advanced our understanding of asteroid composition, size, rotation, regolith spectral and thermal characteristics, surface evolution mechanisms and histories, and dynamical origins. For example, the optical observations of asteroids and spectral measurements of meteorites in the laboratory \citep{nesvorny2005evidence,brunetto2006modeling,vernazza2009solar,cloutis2014establishing,binzel2019compositional,demeo2022connecting,demeo2023isolating} have revealed that most of the population in the inner belt \citep{demeo2014solar} and near-Earth space \citep{binzel2019compositional,sanchez2024population} are Q-, Sq-, or Sq-type. All three types are considered to correspond to ordinary chondrite composition but reflect different degrees of space weathering. Q-type has a fresh (unweathered or refreshed) surface, Sq-type has a moderately weathered surface, and S-type has undergone a higher degree of space weathering \citep{binzel2019compositional,zhang2022diverse}. The term ``space weathering'' refers to the processes that generate regolith and alter the physical and chemical properties of the surfaces of airless bodies by micrometeoroid bombardment and solar wind irradiation \citep{pieters2016space}. For silicate-rich asteroids, the space weathering effects have been summarized as: (1) decreasing spectral reflectance (darkening), (2) moderating spectral absorption features, (3) increasing spectral slope (reddening), and (4) SMFe$^0$ contributes to the first three spectral effects \citep{pieters2016space}. Studying space weathering characteristics and spectral alteration rates is therefore critical to understanding the evolution mechanisms and history of asteroid surfaces \citep{binzel2004observed,nesvorny2005evidence,marchi2005space,marchi2006spectral,vernazza2009solar}.

Additionally, samples collected in situ allow for more in-depth analysis in laboratories. Up to present, three asteroid sample-return missions, namely Hayabusa to 25143 Itokawa \citep{yoshikawa2021hayabusa}, Hayabusa2 to 162173 Ryugu \citep{sei2017hayabusa2}, and OSIRIS-REx to 101955 Bennu \citep{lauretta2017osiris}, have been successfully performed and promoted a comprehensive understanding of asteroids. For example, grains returned from Itokawa provided the first irrefutable evidence linking Sq-type asteroids with LL ordinary chondrites \citep{nakamura2011itokawa,yurimoto2011oxygen}, although Itokawa was estimated as the parent body of LL chondrite via reflectance spectra before the probe returned to Earth \citep{binzel2001muses,abe2006near}. Itokawa has a low content of SMFe$^0$ (\textasciitilde0.05~vol.\%, corresponding to \textasciitilde0.2~wt.\%, \cite{binzel2001muses,hiroi2006developing}) and short surface exposure timescale (10$^3$ to 3 $\times$ 10$^6$ years, \cite{jin2022estimation,nagao2011irradiation}), indicating a moderate space weathering. Samples from Ryugu revealed a CI-chondrite-like composition \citep{greenwood2023oxygen}, short exposure timescale of \textasciitilde5 $\times$ 10$^6$ years \citep{okazaki2022first}, and dehydrated space weathering characteristics \citep{noguchi2023dehydrated}. Analysis of Bennu samples also suggested that its composition is similar to CI chondrites \citep{marty2025noble} and has undergone 2-7 $\times$ 10$^6$ years of space weathering \citep{keller2025space}.

On May 29, 2025, the China National Space Administration successfully launched the Tianwen-2 probe, which plans to perform a two-phase mission: first orbit to explore an Apollo-group asteroid (469219) 2016~HO$_3$ Kamo`oalewa and return its samples in 2027, then orbit to observe an inner main belt active asteroid 311P in 2035. In our previous work \citep{zhang2025tianwen}, we reported the composition and dynamic source region of Kamo`oalewa. To further support the Tianwen-2 mission, in this study, we modeled Kamo`oalewa's shape, rotation period, and pole orientation, and estimated its global distribution of regolith critical size, thermal inertia, regolith thickness, SMFe$^0$ content, and spectral type.
\section{Previous knowledge of Kamo`oalewa}\label{2}
Kamo`oalewa was first discovered on April 27, 2016, by Pan-STARRS at Haleakala Observatory. As the smallest, closest, and most stable quasi-satellite of Earth currently \citep{de2016asteroid}, Kamo`oalewa was named for its oscillating orbit. Currently, there is debate regarding the composition, dynamical origin, thermal inertia, and regolith grain size of Kamo`oalewa. To date, the only visible to near-infrared (VIS-NIR) reflectance spectrum of Kamo`oalewa was obtained by the Large Binocular Telescope and the Lowell Discovery Telescope between 2017 and 2021 \citep{sharkey2021lunar}. Based on the similarity of spectral slope and orbital calculations, \cite{sharkey2021lunar} were the first to propose that Kamo`oalewa originated from the Moon. Subsequently, another orbital calculation work from \cite{castro2023lunar} was consistent with this view. Furthermore, the dynamic simulations from \cite{jiao2024asteroid} suggest that if Kamo`oalewa originated from the Moon, the young lunar impact crater ``Giordano Bruno'' is the most probable source region. However, using the advanced ``seven escape regions orbital dynamics model” \citep{granvik2018debiased} to trace the source region of Kamo`oalewa, \cite{fenucci2021role} proposed that Kamo`oalewa has the highest probability of originating from the inner main belt $\nu_6$ secular resonance.

Here, we briefly introduce a previous work by our team on Kamo'oalewa \citep{zhang2025tianwen}. In that work, we found that highly-energy laser irradiation (which can simulate space weathering process driven by micrometeoroid bombardment \citep{sasaki2001production}) of the LL chondrite powder with a size < 45~$\mu$m could also well reproduce Kamo`oalewa’s observed spectrum (Fig. \ref{figA1}A-D), indicating that Kamo`oalewa is likely to develop an LL-chondrite-compositional, highly space-weathered surface. Meanwhile, using two classic orbital dynamics models, the seven escape regions orbital dynamics model \citep{granvik2018debiased} and the METEODOM model \citep{brovz2024young}, we found that if Kamo`oalewa is an LL chondrite composition, it likely originates from the $\nu_6$ secular resonance, more specifically, the Flora family \citep{zhang2025tianwen}. The $\nu_6$ secular resonance is adjacent to the Flora family, the major source region of NEAs with LL chondrite composition \citep{binzel2019compositional}. Furthermore, we found that Itokawa and 7 objects in the Flora family show a composition similar to that of Kamo`oalewa (Fig. \ref{figA1}E-F). We also noted that laser-irradiated LL chondrite powder is closer to Kamo`oalewa's spectra than the three lunar materials that are used to support the view of Kamo`oalewa's lunar origin (\ref{figA1}C-D). Meanwhile, we noted that the spectrum of the lunar ``Giordano Bruno'' crater does not match that of Kamo`oalewa (Fig. \ref{figA1}E-F), refuting this crater as the origin of Kamo`oalewa. Note that the asteroid source region models used in our early work \citep{zhang2025tianwen} and \cite{fenucci2021role} do not include the Moon, while the idea of the papers \citep{sharkey2021lunar,castro2023lunar,jiao2024asteroid} supporting the lunar origin is to assume that Kamo`oalewa was ejected from the Moon and prove that it could evolve into current orbital state, without considering the main belt or other source regions. Therefore, it is difficult to conclude whether Kamo`oalewa has a higher or lower likelihood of originating from the $\nu_6$ secular resonance (further, the Flora family) or the Moon based on our last work \citep{zhang2025tianwen} and earlier studies \citep{sharkey2021lunar,castro2023lunar,jiao2024asteroid}. However, a recent quantitative estimate \citep{fenucci2026origin} showed that population models of NEAs based on the migration of objects from the main belt can explain the existence of Kamo`oalewa-like objects, without relying on the lunar origin hypothesis. Furthermore, the numerical simulation indicates that the expected number of Kamo`oalewa-like objects originating from the main belt is more than one order of magnitude larger than the expected number of those originating from the Giordano Bruno crater \citep{fenucci2026origin}. In addition, another recent numerical simulation found pathways from the Flora family and the region near the 3:1 mean-motion resonance with Jupiter, to the Earth co-orbital region \citep{wang2026dynamical}. These findings indicate that the dynamic source region of Earth’s quasi-satellites may not be limited to the Moon. Although we did not completely rule out the possibility that other regions on the Moon may be ejection sites for Kamo`oalewa, the better self-consistency of our spectroscopic and orbital dynamics results indicates that Kamo`oalewa is likely an LL-chondrite-compositional asteroid with space weathering and originated from the inner main belt. Please see more information in Appendix \ref{appendixA}.

Based on lightcurve observations, \cite{sharkey2021lunar} found that Kamo`oalewa has a fast rotation period of 28.3$^{+1.8}_{-1.3}$ minutes and is likely less than a hundred meters in size. Subsequently, the shape and regolith size of Kamo`oalewa were discussed by \cite{li2021shape}, who proposed that Kamo`oalewa could retain millimeter- to centimeter-sized regolith grains on the surface. Recently, using a statistical method based on the Yarkovsky-related orbital drift, \cite{liu2024surface} gave a thermal inertia estimation of 402.05$^{+376.29}_{-194.37}$ J m$^{-2}$ K$^{-1}$ s$^{-1/2}$ for Kamo`oalewa, corresponding to a regolith size of millimeters to decimeters. However, another calculation based on the gravity-adhesion-centrifugal force balance showed that the surface of Kamo`oalewa could be covered by regolith grain with a size < 1~cm \citep{ren2024surface}. Additionally, the latest thermal inertia calculation based on observations of the Yarkovsky effect suggested that Kamo`oalewa has a low thermal inertia value of 150$^{+90}_{-45}$ or 181$^{+95}_{-60}$ J m$^{-2}$ K$^{-1}$ s$^{-1/2}$, indicating that its surface is covered by a regolith layer with grain size < 3~mm \citep{fenucci2025astrometry}. Our laser irradiation experiment also revealed that space-weathered LL ordinary chondrite powder with a size < 45~$\mu$m (Fig. \ref{figA1}A-D) rather than the slab (Fig. \ref{figA2}) could well reproduce Kamo`oalewa’s observed reflectance spectrum, implying that the regolith grains on the surface of Kamo`oalewa are fine rather than coarse \citep{zhang2025tianwen}.

To be conservative, the above controversy seems to have no definite conclusion until the Tianwen-2 probe arrives at the asteroid and returns samples. However, in order to respond to these controversies, this study uses multiple methods to update and discuss the contents of the above review. Additionally, since our previous work provided sufficient evidence supporting that Kamo`oalewa develops an LL-chondrite-compositional but ultra-highly space-weathered surface, this study is temporarily based on our previous work \citep{zhang2025tianwen} to extend, aiming to provide a reference for the implementation of the Tianwen-2 mission and give a prediction for Kamo`oalewa.

\section{Methods}\label{3}
\subsection{Shape modeling of Kamo`oalewa}\label{3.1}
We used lightcurve data (2004 03 17.470006 to 2024 03 18.576043) from the IAU Minor Planet Center1\footnote{https://minorplanetcenter.net/} and the Cellinoid method \citep{lu2014cellinoid,zhang2025dcappso} to model the shape, rotation period, and pole orientation of Kamo`oalewa. The Cellinoid model is based on three assumptions: (1) the asteroid shape is composed of eight octants from eight different ellipsoids, divided by six semi-axes a$_1$, a$_2$, b$_1$, b$_2$, c$_1$, and c$_2$, (2) the axis lengths of the asteroid are \textit{$l_a$} $>$ \textit{l$_b$} $>$ \textit{l$_c$}, in which \textit{l$_a$} $=$ \textit{r$_{a1}$} $+$ \textit{r$_{a2}$}, \textit{l$_b$} $=$ \textit{r$_{b1}$} $+$ \textit{r$_{b2}$}, \textit{l$_c$} $=$ \textit{r$_{c1}$} $+$ \textit{r$_{c2}$}, and \textit{r$_{a1}$} to \textit{r$_{c2}$} are the lengths of six semi-axes, respectively, and (3) the asteroid rotates around the principal axis c. Owing to the asymmetric shape of the Cellinoid, the physical parameters of asteroids, such as axis ratios and rotation period, can be fitted more accurately than the traditional three-axis ellipsoid model. The core idea of the Cellinoid method is to first establish a Cellinoid shape, then randomly set the initial values of the axial parameters, rotation period, pole orientation, and phase angle by the program and continuously adjust them as input to generate multiple lightcurves. When generated lightcurves are closest to the actual observed lightcurves, namely mean square error ($\chi$$^2$, calculated method see in equations (5) to (6) in \cite{zhang2025dcappso}) is minimum, the program described in \cite{zhang2025dcappso} will stop running, and the corresponding axial ratios, rotation period, and pole orientation at this time are the most reliable parameters of the asteroid.

For Kamo`oalewa, we set the axis parameters range of the 6 semi-axes from 0 to 1, the rotation period range from 0 to 24~h, the longitude range of pole orientation from 0$^{\circ}$ to 360$^{\circ}$, and the latitude range of pole orientation from $-$90$^{\circ}$ to $+$90$^{\circ}$. When $\chi$$^2$ was minimum, we got the axis ratios of the six semi-axes, rotation period, and pole orientation.

After fixing these parameters, we calculated the lengths of the six semi-axes. Here, an axis length coefficient \textit{c$_0$} is introduced, and the lengths of the six semi-axes can be described by equation (\ref{Eq1}):
\begin{equation}
\left\{
\begin{aligned}
c_0r_{a1}; c_0r_{a2}\\
c_0r_{b1}; c_0r_{b2}\\
c_0r_{c1}; c_0r_{c2}
\end{aligned}
\right. 
\label{Eq1}
\end{equation} Where \textit{c$_0$} was obtained when \textit{L}$_c$($\alpha$) $=$ \textit{L}$_s$($\alpha$), \textit{L}$_c$($\alpha$) refers to the brightness integral of Cellinoid shape at phase angle $\alpha$, \textit{L}$_s$($\alpha$) is the brightness integral of a standard sphere with an effective diameter \textit{D$_{eff}$} at phase angle $\alpha$. The calculation method of brightness integral was described in \cite{lu2014cellinoid}.

For an asteroid, its \textit{D$_{eff}$} could be calculated by the empirical equation (\ref{Eq2}):
\begin{equation}
D_{eff} = \frac{1329}{\sqrt{Pv}}10^{-0.2H}
\label{Eq2}
\end{equation} where \textit{Pv} is the visual geometric albedo, and \textit{H} is the absolute magnitude.

For Kamo`oalewa, we used the \textit{H} $=$ 24.33 mag \citep{sharkey2021lunar} and the \textit{Pv} $=$ 0.1; thereby, the calculated \textit{D$_{eff}$} is 0.057~km. \textit{Pv} $=$ 0.1 is chosen here because the reflectance at 0.55~$\mu$m (the reflectance at 0.55~$\mu$m of meteorites is generally close to the visible geometric albedo of the asteroid) of our space-weathered LL chondrite powder (whose spectrum best matches Kamo`oalewa) is close to 0.1 (see Fig. \ref{figA1}B in Appendix \ref{appendixA}).
\subsection{Estimation of the global distribution of regolith critical size of Kamo`oalewa}\label{3.2}
To estimate the critical (maximum) size (diameter) of regolith grains at the surface of the Kamo`oalewa, we considered three main forces: gravity, cohesive force, and centrifugal force.

In the first step, we estimated the gravity experienced by the regolith grains. For a regular ellipsoid asteroid (regular ellipsoid means that \textit{r$_{a1}$} $=$ \textit{r$_{a2}$}, \textit{r$_{b1}$} $=$ \textit{r$_{b2}$, \textit{r$_{c1}$} $=$ \textit{r$_{c2}$}}), \textit{V$_e$} at point (\textit{x$_i$}, \textit{y$_i$}, \textit{z$_i$}) outside the body can be described by equation (\ref{Eq3}) \citep{scheeres2010scaling}:
\begin{equation}
\begin{split}
& V_e = -\int_{0}^{2\pi}\int_{0}^{\pi}\int_{0}^{1} \\ & \frac{G\rho{l_a}{l_b}{l_c}r\sin{\varphi}\mathrm{d}r\mathrm{d}\varphi\mathrm{d}\theta}{8\sqrt{(\frac{{l_a}}{2}\cos{\theta}\sin{\varphi} - {x_i})^2 + (\frac{{l_b}}{2}\sin{\theta}\sin{\varphi} - {y_i})^2 + (\frac{{l_c}}{2}\cos{\varphi} - {z_i})^2}}
\end{split}
\label{Eq3}
\end{equation} where \textit{G} $=$ 6.674 $\times$ 10$^{-11}$ m$^3$ kg$^{-1}$ s$^{-2}$ is the gravitational constant, $\rho$ is the bulk density of an asteroid, \textit{r} $=$ $\sqrt{{x_i}^2 + {y_i}^2 + {z_i}^2}$ is the radial distance, $\varphi$ is the polar angle, and $\theta$ is the azimuth angle in the spherical coordinate system. \textit{l$_a$}, \textit{l$_b$}, and \textit{l$_c$} are the lengths of the three axes a, b, and c, respectively.

Since our Cellinoid shape of Kamo`oalewa is an irregular ellipsoid (irregular ellipsoid means that \textit{r$_{a1}$} $\neq$ \textit{r$_{a2}$}, \textit{r$_{b1}$} $\neq$ \textit{r$_{b2}$}, \textit{r$_{c1}$} $\neq$ \textit{r$_{c2}$}, composed of eight octants of eight regular ellipsoids, its gravitational potential \textit{V$_c$} was described as the integral of eight octants, as shown in equation (\ref{Eq4}):
\begin{equation}
V_c = \sum_{n=1}^8\iiint_{\Omega_n}f_n\mathrm{d}\sigma
\label{Eq4}
\end{equation} where $\Omega_n$ is the \textit{n}-th part of the Cellinoid shape, $f_n$ is the n-th integrand, and $f_n$ can be described as equation (\ref{Eq5}):
\begin{equation}
\begin{split}
& f_n = \\ & \frac{G\rho{l_{am}}{l_{bp}}{l_{cq}}r\sin{\varphi}}{8\sqrt{(\frac{{l_{am}}}{2}\cos{\theta}\sin{\varphi} - {x_i})^2 + (\frac{{l_{bp}}}{2}\sin{\theta}\sin{\varphi} - {y_i})^2 + (\frac{{l_{cq}}}{2}\cos{\varphi} - {z_i})^2}}
\label{Eq5}
\end{split}
\end{equation} where \textit{m}, \textit{p}, \textit{q} $=$ 1, 2. For example, the first part of the Cellinoid is implemented when \textit{m} $=$ 1, \textit{p} $=$ 1, and \textit{q} $=$ 1, namely the ellipsoid a$_1$, b$_1$, c$_1$ in the region \textit{r} > 0, $\pi$ > $\varphi$ > $\pi$⁄2, $\pi$⁄2 > $\theta$ > 0. We used $\rho$ $=$ 3200 kg m$^{-3}$ (the average bulk density of LL chondrites \citep{macke2010survey}) for Kamo`oalewa because we considered that Kamo`oalewa is a single boulder (non-rubble pile) structure due to its fast rotation \citep{sharkey2021lunar}.

Then, the gravitational acceleration  \textit{a$_g$} can be determined by computing the partial derivatives of the gravitational potential in each direction, as shown in equation (\ref{Eq6}):
\begin{equation}
a_g = -\nabla{V_c} = \left[\frac{\partial {V_c}}{\partial x}\frac{\partial {V_c}}{\partial y}\frac{\partial {V_c}}{\partial z}\right]
\label{Eq6}
\end{equation}

The gravity \textit{F$_g$} can be determined as equation (\ref{Eq7}):
\begin{equation}
F_g  = ma_g
\label{Eq7}
\end{equation} where \textit{m} is the mass of the regolith grains, which can be described as equation (\ref{Eq8}):
\begin{equation}
m  = \rho \frac{4}{3} \pi R^3
\label{Eq8}
\end{equation} where $\rho$ $=$ 3200 kg m$^{-3}$ is the bulk density of the regolith single grain and \textit{R} is the radius of the regolith grain. It is worth noting that the direction of gravity depends on the gravitational potential and does not usually point toward the core of the ellipsoid.

In the second step, we estimated the cohesive force \textit{F$_s$} between the regolith grains and Kamo`oalewa’s body. The traditional view, based on empirical models, believed that adhesion would continue to increase with increasing particle size. However, \cite{nagaashi2023high} accurately measured the adhesion of different meteorites in the laboratory for the first time and performed microscopic analysis, finding that adhesion was related to the number of contact points between the sample and other substances. Through experiments and calculations, \cite{nagaashi2023high} gave new adhesion results and successfully explained the phenomenon that grains on the surface of asteroids are more likely to slide and migrate. That is, early models overestimated the adhesive force and were unable to explain the high mobility of the asteroid surface. Therefore, we adopted the cohesive force from \cite{nagaashi2023high} rather than that from the traditional model. \cite{nagaashi2023high} found that the cohesive force of grains or rocks smaller than \textasciitilde0.7~m can be regarded as constants. For LL3.5 chondrite grains, the cohesive force ranges from 39 to 300~nN; for LL5 chondrite grains, the cohesive force ranges from 69 to 330~nN; for LL6 chondrite grains, the cohesive force ranges from 114 to 510~nN \citep{nagaashi2023high}. Given the small size of Kamo`oalewa, it is assumed that the size of its regolith is no larger than 0.7~m, and therefore, 300~nN is taken as a typical value in this study.

In the third step, we estimated the centrifugal force experienced by the regolith grains. As with gravity, the direction of the centrifugal force should also be determined before estimating the critical size of grains on the surface of Kamo`oalewa. Since Kamo`oalewa was assumed to rotate around the c-axis, the centrifugal force vector is radially outward from the c-axis and perpendicular to the c-axis. The centrifugal force at point (\textit{x$_i$}, \textit{y$_i$}, \textit{z$_i$}) can be described as equation (\ref{Eq9}):
\begin{equation}
F_c = m \frac{4\pi^2}{T^2} r_p
\label{Eq9}
\end{equation} where \textit{T} is the rotation period. \textit{r$_p$} $=$ $\sqrt{{x_i}^2 + {y_i}^2 + {z_i}^2}$ is the distance between the grain position and the geometric center of the asteroid.

In the last step, we estimated the critical size of the regolith grains by centrifugal force $\geq$ the sum of gravity and cohesive force.
\subsection{Thermal parameters measurement of laser-irradiated LL chondrite powder}\label{3.3}
In the temperature range of 253.15 to 473.15~K, we measured the bulk density, thermal diffusivity, and specific heat capacity of laser-irradiated LL chondrite powder.

In the first step, we placed the uncompacted powder into a 10~mm $\times$ 10~mm $\times$ 3~mm transparent sapphire sample cell and weighed its mass. Thus, we obtained the bulk density (which contains porosity) of the powder, which is $\rho_p$ $=$ 700 kg m$^{-3}$.

In the second step, we measured the thermal diffusivity of the powder in the sapphire sample cell using the laser flash method. The principle of this method is to irradiate a homogeneous sample of a specific size with a high-intensity, short-time energy pulse laser at a set temperature \textit{T$_p$}. The lower surface of the sample absorbs energy to instantly increase the temperature, and as the hot end, it transfers energy to the cold end (upper surface) in an ideal one-dimensional heat conduction manner. Record the temperature rise curve at the center of the sample’s upper surface, that is, the half temperature rise curve. Then, the thermal diffusivity $\alpha$ was calculated by recording the half-temperature rise curve.

In the third step, differential scanning calorimetry (DSC) was used to measure the constant pressure specific heat capacity (\textit{C$_P$}) of the powder. Thus, the thermal conductivity (\textit{K$_p$}) of the powder could be calculated by equation (\ref{Eq10}):
\begin{equation}
K_P = \alpha C_P \rho_p
\label{Eq10}
\end{equation}

Finally, the thermal inertia (\textit{$\Gamma_p$}) of the powder could be calculated by equation (\ref{Eq11}):
\begin{equation}
\Gamma_p = \sqrt{\rho_p C_P K_P}
\label{Eq11}
\end{equation}
\subsection{Estimation of thermal inertia and regolith thickness of Kamo`oalewa}\label{3.4}
Currently, Kamo`oalewa’s thermal inertia and regolith thickness are not very clear due to the lack of direct thermal observations. However, the laser-irradiated LL chondrite powder has shown a spectrum matching Kamo`oalewa, so we believe that the powder’s thermal inertia can approximately reflect the situation of Kamo`oalewa. Meanwhile, the regolith thickness of Kamo`oalewa can be simply estimated using equation (\ref{Eq12}):
\begin{equation}
l_s = \sqrt{\frac{K_p}{\rho_p C_P w_a}}
\label{Eq12}
\end{equation} where \textit{l$_s$} is the thermal skin depth, reflecting the depth of an asteroid’s regolith affected by the thermal cycles and can represent the lower limit of the regolith thickness. \textit{K$_p$} is the thermal conductivity, \textit{C$_P$} is the specific heat capacity, and \textit{w$_a$} is the asteroid’s rotation angular velocity.
\subsection{Estimation of SMFe$^0$ content of Kamo`oalewa’s regolith} \label{3.5}
We used an improved “radiative transfer mixing model” \citep{lawrence2007radiative} to estimate the SMFe$^0$ content in the regolith of Kamo`oalewa. This model considers the contributions of troilite and iron-nickel metal to the spectrum, making it more applicable to silicate-rich asteroids.

In the first step, we modeled the reflectance spectrum of fresh (containing 0 wt.\% SMFe$^0$) LL chondrite (Kheneg Ljouâd) powder (size < 45~$\mu$m, average size is 28.82~$\mu$m, measured by laser particle size analyzer) as the initial state (without space weathering) of Kamo`oalewa. Assuming that the asteroid regolith particles are an intimate mixture and their particle size is much larger than the wavelength, the bidirectional reflectance (\textit{r$_c$}) of asteroidal regolith grains can be expressed as equation (\ref{Eq13}) \citep{hapke1981bidirectional}:
\begin{equation}
\begin{split}
r_c =& \frac{w_{avg}}{4\pi}\frac{1}{\mu_0 + \mu}\{[1 + B(g)]P(g) \\
& + H(\mu_0)H(\mu) - 1\}
\label{Eq13}
\end{split}
\end{equation} where $\mu$ = $\cos(\textit{e})$, $\mu_0$ = $\cos(\textit{i})$. \textit{e} and \textit{i} are the emission angle and incidence angle, respectively, and \textit{g} is the phase angle. In this study, we used \textit{e} = 0$^\circ$, \textit{i} = 30$^\circ$, and \textit{g} = 30$^\circ$.

In equation (\ref{Eq13}), the average single scattering albedo (\textit{w$_{avg}$}) of the grains can be expressed as equation (\ref{Eq14}):
\begin{equation}
w_{avg} = \frac{\sum_{j}\frac{w_jV_j}{D_j} }{\sum_{j}\frac{V_j}{D_j}}
\label{Eq14}
\end{equation} where \textit{w$_j$} is the single scattering albedo of each major mineral endmember grain (including olivine, orthopyroxene, clinopyroxene, plagioclase, troilite, and iron-nickel metal in this study), \textit{V$_j$} is the volume fraction of each mineral endmember, and \textit{D$_j$} is the average diameter of grains. In this study, the \textit{V$_j$} of various minerals are as follows: olivine is 57.73 vol.\%, orthopyroxene is 19.87 vol.\%, clinopyroxene is 5.61 vol.\%, plagioclase is 12.23 vol.\%, troilite is 2.67 vol.\%, and iron-nickel metal is 0.56 vol.\%. These values are derived from experimental measurements of the mineral abundances in Kheneg Ljouâd (see Fig. \ref{figB1}). Additionally, the \textit{D$_j$} is 28.82~$\mu$m.

For silicate mineral grains such as olivine, orthopyroxene, clinopyroxene, and plagioclase, their \textit{w$_j$} can be expressed as equation (\ref{Eq15}) \citep{hapke2001space}:
\begin{equation}
w_j = S_e + (1 - S_e ) \frac{1-S_i}{1-S_i\Theta}\Theta
\label{Eq15}
\end{equation} where \textit{S$_e$} represents the surface reflection coefficient for externally incident light, \textit{S$_i$} denotes the reflection coefficient for internally scattered light, and $\Theta$ is the single-pass transmission of the grain.

For silicate mineral grains (imaginary refractive index $\ll$ 1), \textit{S$_e$} can be approximated by equation (\ref{Eq16}) \citep{hapke2001space}:
\begin{equation}
S_e = \frac{(n-1)^2}{(n+1)^2} + 0.05
\label{Eq16}
\end{equation} where \textit{n} is the real index of refraction, which can be derived from equations (\ref{Eq5}) to (\ref{Eq8}) \citep{lucey1998model,warell2010hapke}:

for olivine: 
\begin{equation}
n_{ol} = 1.827 - 0.192 (Mg\#)
\label{Eq17}
\end{equation}

for orthopyroxene: 
\begin{equation}
n_{opx} = 1.768 - 0.118 (Mg\#)
\label{Eq18}
\end{equation}

for clinopyroxene: 
\begin{equation}
n_{cpx} = 1.726 - 0.082 (Mg\#)
\label{Eq19}
\end{equation}

for plagioclase: 
\begin{equation}
n_{pl} = 1.523 + 0.0227 (An\#) +0.0264 (An\#)^2
\label{Eq20}
\end{equation} where the magnesium number \textit{Mg\#} is Mg/(Mg+Fe), and anorthite number \textit{An\#} is Ca/(Ca+Na). For the Kheneg Ljouâd chondrite, the \textit{Mg\#} in olivine is 0.69, the \textit{Mg\#} in orthopyroxene is 0.729, the \textit{Mg\#} in clinopyroxene is 0.462, and the \textit{An\#} in plagioclase is 0.106 (see Appendix \ref{appendixB}).

For silicate mineral grains, \textit{S$_i$} and $\Theta$ can be expressed as equations (\ref{Eq21}) \citep{lucey1998model} and (\ref{Eq22}) \citep{hapke2001space}:
\begin{equation}
S_i = 1.014 - \frac{4}{n(n+1)^2}
\label{Eq21}
\end{equation}
\begin{equation}
\Theta = \mathrm{e}^{-{\alpha_w}	\left\langle D \right \rangle}
\label{Eq22}
\end{equation} where $\left\langle D \right \rangle$ is the mean optical path length of the grains, and $\alpha_w$ is the absorption coefficient of the medium. They can be expressed as equations (\ref{Eq23}) \citep{hapke2012theory} and (\ref{Eq24}) \citep{lucey2011optical}:
\begin{equation}
\left\langle D \right \rangle = \frac{2}{3}	\left[n^2 - \frac{1}{n}(n^2-1)^{\frac{3}{2}}\right]D
\label{Eq23}
\end{equation}
\begin{equation}
\alpha_w = \alpha_h + \alpha_{Fe} + \alpha_g
\label{Eq24}
\end{equation} where \textit{D} is the grain diameter (28.82~$\mu$m used in this study). \textit{$\alpha_h$} is the absorption coefficient of the host material (olivine, orthopyroxene, clinopyroxene, and plagioclase) \citep{lucey2011optical}:
\begin{equation}
\alpha_h = \frac{4 \pi n_h k_h}{\lambda}
\label{Eq25}
\end{equation} where \textit{n$_h$} is the real index of refraction of the host material, calculated using equations (\ref{Eq17}) to (\ref{Eq20}). \textit{k$_h$} represents the coefficient of the imaginary index of refraction for the host material, and $\lambda$ denotes the wavelength. The \textit{k$_h$} values for olivine, orthopyroxene, and clinopyroxene in the 0.6--2.2~$\mu$m region were derived using the MGM method \citep{trang2013near}, while for the 0.5--0.6~$\mu$m region, equations (\ref{Eq13}) to (\ref{Eq20}) in \cite{lucey1998model} were adopted because the MGM method does not work well in featureless regions. The standard reflectance spectra used were from the USGS Spectral Library (Olivine KI3291, Bronzite HS9, Augite WS592). The \textit{k$_h$} value for plagioclase was derived using a standard reflectance spectrum (Anorthite HS201) from the USGS Denver spectral library and equations (\ref{Eq13}) to (\ref{Eq20}) \citep{lucey1998model}.

In equation (\ref{Eq24}), \textit{$\alpha_{Fe}$} is the absorption coefficient of the SMFe$^0$ coating on the grains, which can be calculated according to equation (\ref{Eq26}) \citep{lucey2011optical}:
\begin{equation}
\alpha_{Fe} = \frac{36\pi z f_c \rho_h}{\lambda\rho_{Fe}}
\label{Eq26}
\end{equation} where \textit{f$_c$} is the mass fraction of SMFe$^0$ coating the entire host mineral grains (for fresh LL chondrite powder, \textit{f$_c$}  = 0). \textit{$\rho_h$} is the density of the host material grains (3320 kg m$^{-3}$ for olivine, 3550 kg m$^{-3}$ for orthopyroxene, 3400 kg m$^{-3}$ for clinopyroxene is, and 2680 kg m$^{-3}$ for plagioclase). \textit{$\rho_{Fe}$} is the density of iron, 7870 kg m$^{-3}$. \textit{z} can be described as equation (\ref{Eq27}) \citep{lucey2011optical}:
\begin{equation}
z = \frac {{n_h}^3 n_{Fe} k_{Fe}}{{({n_{Fe}}^2 - {k_{Fe}}^2 + 2{n_h}^2)}^2 + {(2 {n_{Fe}} {k_{Fe}})}^2}
\label{Eq27}
\end{equation} where \textit{$n_{Fe}$} is the real index of refraction of SMFe$^0$ and \textit{$k_{Fe}$} is the coefficient of the imaginary index of refraction of SMFe$^0$. These values were obtained from \cite{cahill2019optical}.

In equation (\ref{Eq24}), \textit{$\alpha_g$} is the absorption coefficient of SMFe$^0$ in the host materials, which can be described as equation (\ref{Eq28}) \citep{lucey2011optical}:
\begin{equation}
\alpha_g = \frac{3 q_a f_g \rho_h}{2 d_{Fe} \rho_{Fe}}
\label{Eq28}
\end{equation} where \textit{$q_a$} is the absorption efficiency of an iron particle, \textit{$f_g$} is the mass fraction of SMFe$^0$ in the host materials, and \textit{$d_{Fe}$} is the diameter of the SMFe$^0$ particles in the host materials (200~nm was used in this study).

Thus, the mass fraction of the total SMFe$^0$ (\textit{f}) in regolith can be described as:
\begin{equation}
f = f_c + f_g
\label{Eq29}
\end{equation}

For two metallic minerals (troilite and iron-nickel), we used equations (\ref{Eq30}) to (\ref{Eq32}) \citep{lawrence2007radiative} to calculate their \textit{$w_j$}:
\begin{equation}
w_j = \frac{w_p + w_s}{2}
\label{Eq30}
\end{equation}
\begin{equation}
w_p = \frac{(n^2 + k^2)\cos^2{(\varphi_0)} - 2n\cos{(\varphi_0)} + 1}{(n^2 + k^2)\cos^2{(\varphi_0)} + 2n\cos{(\varphi_0)} + 1}
\label{Eq31}
\end{equation}
\begin{equation}
w_s = \frac{n^2 + k^2 - 2n\cos{(\varphi_0)} + \cos^2{(\varphi_0)}}{n^2 + k^2 + 2n\cos{(\varphi_0)} + \cos^2{(\varphi_0)}}
\label{Eq32}
\end{equation} where \textit{$\varphi_0$} is the incidence angle (30$^\circ$ in this study). \textit{n} is the real index of refraction, and \textit{k} is the coefficient of the imaginary index of refraction. Values of \textit{n} and \textit{k} for troilite were cited from \cite{egan1977rings}. Values of \textit{n} and \textit{k} for iron-nickel metal were approximately replaced by that of nickel \citep{cahill2019optical} because the atomic ratio of iron to nickel in Kheneg Ljouâd’s tetrataenite is about 40:60 (see Appendix \ref{appendixB}).

In equation (\ref{Eq13}), the backscatter function \textit{B(g)} can be expressed as equation (\ref{Eq33}) \citep{hapke2012theory}:
\begin{equation}
B (g) = \frac{B_0}{1 + \left(\frac{1}{h}\right) \tan{\left(\frac{g}{2}\right)}}
\label{Eq33}
\end{equation} where \textit{$B_0$} is the amplitude of the opposition effect (set to 1 in this study). \textit{g} is the phase angle. \textit{h} is the angular width parameter of the opposition effect, which can be expressed as equation (\ref{Eq34}) \citep{hapke2012theory}: 
\begin{equation}
h = \frac{3\sqrt{3}}{8} \frac{K\phi}{\ln{\left(\frac{a_l}{a_s}\right)}}
\label{Eq34}
\end{equation} where \textit{$a_l$} and \textit{$a_s$} are the maximum and minimum sizes, respectively (in this study, their values are 45~$\mu$m and 0.35~$\mu$m, measured on fresh powder of Kheneg Ljouâd by a laser particle size analyzer). $\phi$ is the filling factor, which is set to 0.41 in this study. \textit{K} is the porosity coefficient, which can be expressed as equation (\ref{Eq35}) \citep{hapke2012theory}:
\begin{equation}
K = \frac{\ln{\left( 1 - 1.209 \phi^{\frac{2}{3}} \right)}}{1.209 \phi^ \frac{2}{3}}
\label{Eq35}
\end{equation}

In equation (\ref{Eq13}), the single-particle phase function \textit{P(g)} can be expressed as equation (\ref{Eq36}) \citep{hapke2012theory}:
\begin{equation}
P(g) = 1 + b\cos{(g)} + c[1.5 \cos^2{(g)} -0.5]
\label{Eq36}
\end{equation}

Here, \textit{b} is -0.4, and \textit{c} is 0.25. In equation (\ref{Eq13}), the isotropic scattering function \textit{H($\mu$)} can be expressed as equation (\ref{Eq37}) \citep{hapke2012theory}:
\begin{equation}
\begin{split}
H(\mu) = &  \Bigg\{1 - \left( 1 - \sqrt{1 - w_{avg}} \right) \mu \\
& \left[ r_0 + ( 1 - \frac{r_0}{2} - r_0 \mu ) \ln{\frac{1 + \mu}{\mu}} \right]\Bigg\}^{-1}  
\end{split}
\label{Eq37}
\end{equation} where \textit{$r_0$} could be described as equation (\ref{Eq38}):
\begin{equation}
r_0 = \frac{2}{1 + \sqrt{1 - w_{avg}}} -1
\label{Eq38}
\end{equation}

Thus, when parameter \textit{f} in equation (\ref{Eq29}) is 0, using equations (\ref{Eq13}) to (\ref{Eq38}), we modeled the reflectance spectrum of fresh Kheneg Ljouâd’s powder.

In the second step, we continue to increase the value of \textit{f} in equation (\ref{Eq29}) until the modeled spectral curve matches well with that of Kamo`oalewa. At that time, the corresponding \textit{f} value is the SMFe$^0$ content in the regolith of Kamo`oalewa that we estimated.

Before the estimation, to test the effectiveness of the model, we calculated the average SMFe$^0$ content on the surface of Itokawa. For Itokawa, the used mineral abundance in this study is 64 vol.\% olivine, 19 vol.\% orthopyroxene, 3 vol.\% clinopyroxene, 11 vol.\% plagioclase, 2 vol.\% troilite, and 0.02 vol.\% iron-nickel metal. These data are cited from the analysis of the returned sample of Itokawa \citep{nagao2011irradiation}. The other parameters, such as grain diameter, grain density, \textit{Mg\#}, \textit{An\#}, \textit{n}, and \textit{k} are the same as those estimated for the SMFe$^0$ content of Kamo`oalewa because both have an LL chondrite composition. As a result, Itokawa’s average SMFe$^0$ content was best estimated to be 0.2 wt.\% (Fig. \ref{figC1} in Appendix \ref{appendixC}). This is consistent with previous estimates of ~0.05 vol.\% \citep{binzel2001muses,hiroi2006developing}, indicating that our model can be well applied to Kamo`oalewa.
\subsection{Detection of spectral type} \label{3.6}
The spectral data of Kamo`oalewa is discontinuous and has less information in the near-infrared region, comparing it to Bus-DeMeo standard spectral types and performing a chi-square test, the error will be quite large. If only use the visible range data for classification, we cannot get very good results, because the spectral curves of most asteroids in the visible band are very similar (this is why \cite{demeo2009extension} expanded the data used in the classification scheme to the near-infrared region), no good reference value. Nevertheless, the laser-irradiated LL chondrite powder allows us to obtain a virtual spectral type as a substitute for Kamo`oalewa because the irradiated LL chondrite powder showed a spectrum that matched well with that of Kamo`oalewa. For the reflectance spectrum of irradiated powder, an asteroid classification online tool from the MIT website\footnote{http://smass.mit.edu/busdemeoclass.html} was used to calculate the principal component parameters and spectral type defined in the Bus-DeMeo taxonomy system \citep{demeo2009extension}.
\section{Results}\label{4}
\subsection{Shape, rotation period, and pole orientation of Kamo`oalewa}\label{4.1}
\begin{figure}
   \includegraphics[width=9 cm]{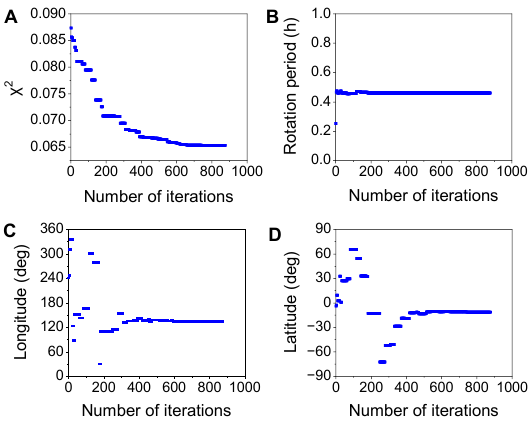}
     \caption{Modeled parameters change with the number of iterations. (A) Mean square error $\chi^2$ changes with the number of iterations. (B) Rotation period changes with the number of iterations. (C) Longitude changes with the number of iterations. (D) Latitude changes with the number of iterations. When running the 877th iteration, $\chi^2$ is minimum, and the rotation period, longitude, and latitude converge to 0.461~h, 134.7$^{\circ}$, and $-$11.4$^{\circ}$, respectively.}
     \label{fig1}
     \end{figure}

Using the method in Section \ref{3.1} and running 877 iterations, we obtained the minimum $\chi^2$ (Fig. \ref{fig1}A). At this time, the corresponding rotation period is 0.461~h (27.66~min), with pole orientation at 134.7$^{\circ}$ longitude and -11.4$^{\circ}$ latitude in ecliptic coordinates, as shown in Fig. \ref{fig1}B--D. The corresponding axis parameters are a$_1$ = 0.999, a$_2$ = 0.537, b$_1$ = 0.533, b$_2$ = 0.496, c$_1$ = 0.562, and c$_2$ = 0.314. When normalized the axis parameter of a$_1$ to 1, the axis ratios could be obtained. After fixing these parameters, we obtained the axis length coefficient of 44.386. Thereby, the lengths of six semi-axes are \textit{r$_{a1}$} = 44.322~m, \textit{r$_{a2}$}  = 23.830 m, \textit{r$_{b1}$}  = 23.651 m, \textit{r$_{b2}$}  = 22.007 m, \textit{r$_{c1}$}  = 24.957 m, and \textit{r$_{c2}$}  = 13.928 m. Fig. \ref{fig2} shows the shape and pole orientation viewed from different directions.

\subsection{Global distribution of regolith critical size of Kamo`oalewa}\label{4.2}
\begin{figure*}
    \centering
   \includegraphics[width=18 cm]{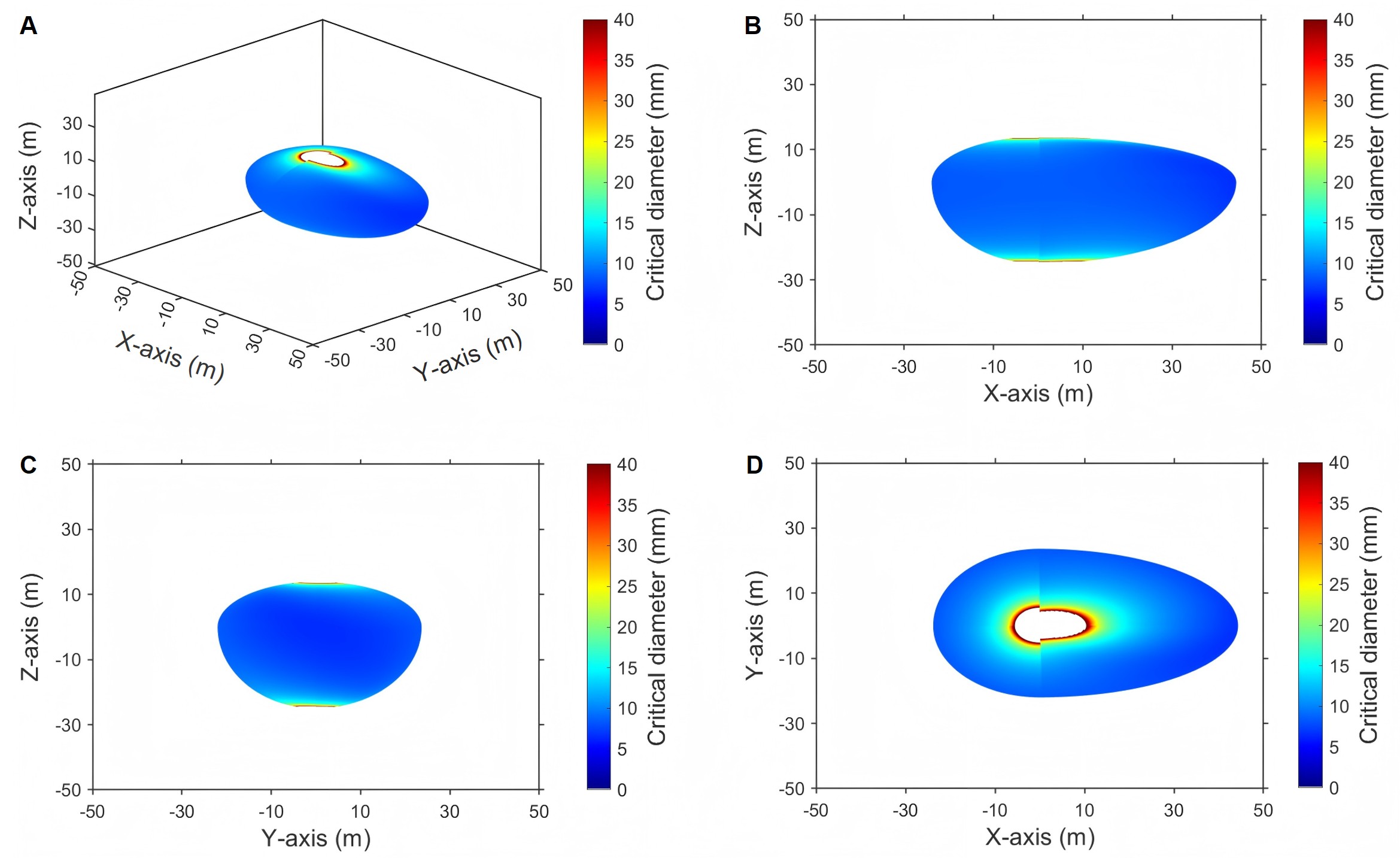}
     \caption{Global distribution of regolith critical size of modeled Kamo`oalewa shape viewed from different views (A--D). Kamo`oalewa’s shape was modeled as 68.15~m $\times$ 45.66~m $\times$ 38.89~m, and the pole orientation was 134.7$^{\circ}$ in longitude and $-$11.4$^{\circ}$ in latitude (ecliptic coordinate system). The blue to red colors represent the increasing critical size of the regolith grains. Regolith grains with a critical size < 2~cm can stably remain on 93.83\% of the global surface area of Kamo`oalewa, of which grains < 1~cm in diameter occupy 67.16\% of global surface area. Two white zones that appear around poles (occupy 3.4\% of the global surface area) could remain the grains with a critical size > 4~cm.}
     \label{fig2}
     \end{figure*}

Using the method in Section \ref{3.2}, we obtained a global distribution of regolith critical size for the modeled shape of Kamo`oalewa. As shown in Fig. \ref{fig2}, regolith grains with a critical size < 2~cm can remain stable over 93.83\% of the global surface area of Kamo`oalewa, in which grains < 1~cm in diameter occupy 67.16\% of global surface area, whereas grains with a critical size > 4~cm can be retained in regions around two poles (white regions in Fig. \ref{fig2}, occupy 3.4\% of the global surface area). This result indicates that most of the surface area of Kamo`oalewa is dominated by fine-grained regolith, which is consistent with what we found that fine-sized space-weathered LL chondrite powder can match the observed spectrum of Kamo`oalewa (see Fig. \ref{figA1}).

\subsection{Thermal conductivity and thermal inertia of laser-irradiated LL chondrite powder}\label{4.3}
\begin{figure*}
    \centering
   \includegraphics[width=18 cm]{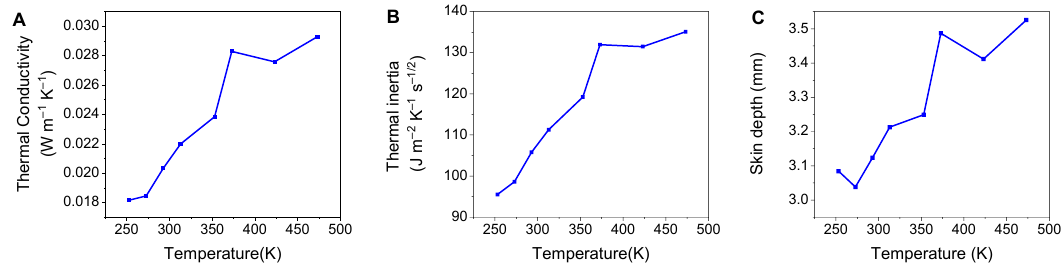}
     \caption{Thermal parameters of space-weathered LL chondrite powder and skin depth of the regolith of Kamo`oalewa. (A) Thermal conductivity changes with temperature. (B) Thermal inertia changes with the temperature. (C) In the temperature range of 253.15 to 473.15~K, the skin depth ranges from 3.04 to 3.53~mm.}
     \label{fig3}
     \end{figure*}

Using the method in Section \ref{3.3}, we obtained the curves of thermal conductivity and thermal inertia of space-weathered LL chondrite powder changing with temperature. As shown in Fig. \ref{fig3}, the sample shows low thermal conductivity and thermal inertia. In the temperature range of 253.15 to 473.15~K, the thermal conductivity increases from 0.018 to 0.029 W m$^{-1}$ K$^{-1}$ (Fig. \ref{fig3}A), and thermal inertia increases from 95.52 to 135.09~J m$^{-2}$ K$^{-1}$ s$^{-1/2}$ (Fig. \ref{fig3}B).

\subsection{Thermal inertia and regolith thickness of Kamo`oalewa}\label{4.4}
As described in Section \ref{3.4}, we approximate the thermal inertia of laser-irradiated LL chondrite powder as a substitute for the value of Kamo`oalewa, which is 95.52 to 135.09~J m$^{-2}$ K$^{-1}$ s$^{-1/2}$. Meanwhile, using the rotation period of Kamo`oalewa that we calculated and the equation (\ref{Eq12}), we obtained the ``skin depth” is 3.04 to 3.53~mm (corresponding to 253.15 to 473.15~K, see Fig. \ref{fig3}C). This means that Kamo`oalewa may have developed a regolith at least 3~mm thick.

\begin{figure}
   \includegraphics[width=9cm]{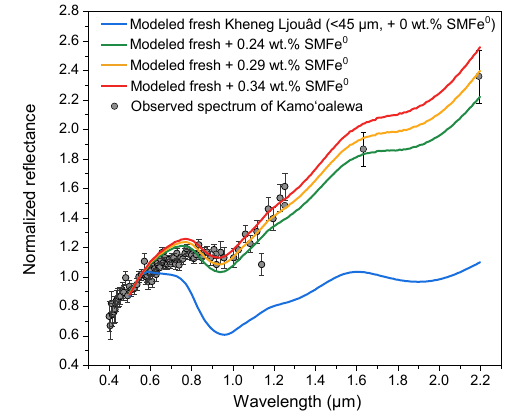}
     \caption{Estimation of SMFe$^0$ content in the regolith of Kamo`oalewa. The spectrum of Kamo`oalewa was best matched to a model of fresh Kheneg Ljouâd (a LL5/6 chondrite) powder with 0.29 $\pm$ 0.05 wt.\% SMFe$^0$. Spectra are normalized at 0.55~$\mu$m.}
     \label{fig4}
     \end{figure}
\subsection{SMFe$^0$ content of Kamo`oalewa’s regolith}\label{4.5} 
Using the method in Section \ref{3.5}, we obtained the estimated SMFe$^0$ content in the regolith of Kamo`oalewa. As shown in Fig. \ref{fig4}, the modeled initial spectrum of Kamo`oalewa (modeled spectrum of fresh Kheneg Ljouâd powder) shows a deep absorption at 1~$\mu$m and a reddish spectral slope. When the SMFe$^0$ content was added to 0.29 $\pm$ 0.05~wt.\%, the corresponding spectral curves best matched that of Kamo`oalewa (Fig. \ref{fig4}). This is marginally higher than the SMFe$^0$ content of Itokawa (0.2 wt.\%) and is consistent with that Kamo`oalewa developed a highly space-weathered surface.
\subsection{Potential spectral type of Kamo`oalewa}\label{4.6}
Using the method in Section \ref{3.6}, we obtain \textit{PC1’} (first principal component) as $-$0.3833, \textit{PC2’} (second principal component) as 0.0896, and the spectral slope as 0.9145. The online classification tool further suggested the classification as Sqw-type, a non-standard type defined in \cite{demeo2009extension}. The term ``-w'' means ``weathered'' and generally with red spectral slope, for example, Sw refers to space-weathered S-type and Sqw-type refers to space-weathered Sqw-type \citep{demeo2009extension}. \cite{demeo2009extension} classified 11 asteroids into Sqw-type (such as (39) Laetitia, (512) Taurinensis, (1329) Eliane, (1807) Slovakia, (2064) Thomsen, (3102) Krok, (3122) Florence, (3198) Wallonia, (3920) Aubignan, (4407) Taihaku, and (6239) Minos), which were generally classified as S-type in the Bus taxonomy system. Therefore, although Sqw is not a standard type, it does reflect the presence of ``space weathering'' on Kamo`oalewa. This is consistent with Kamo`oalewa having a higher SMFe$^0$ content and space weathering degree than Sq-type Itokawa.  Please note that our Sqw-type is the result of the Bus-DeMeo taxonomy system \citep{demeo2009extension}. If the Bus taxonomy system \citep{bus2002phaseb,bus2002phasea} is used, this result may become S-type because this system just used the VIS region data and did not define the ``-w'' term. However, these do not affect the conclusion that Kamo`oalewa is a highly space-weathered asteroid.

\cite{binzel2019compositional} proposed a space weathering parameter ($\Delta\eta$) to evaluate the space weathering degree of the S-complex asteroids. This parameter was calculated by equation (\ref{Eq39}):
\begin{equation}
\Delta\eta = \frac{-\frac{1}{3}PC2' + PC1' + 0.5}{1.0541}
\label{Eq39}
\end{equation}

Based on statistics of asteroids, \cite{binzel2019compositional} roughly divided asteroids into three species: fresh or unweathered ($\Delta\eta$ < \textasciitilde0.2), intermediate space weathering (\textasciitilde0.2 < $\Delta\eta$ < \textasciitilde0.48), and saturated space weathering ($\Delta\eta$ > \textasciitilde0.48). Using equation (\ref{Eq39}), we also calculated the $\Delta\eta$ for irradiated LL chondrite powder. However, its calculated value was 0.0823, falling into an unweathered range. This is very strange because Itokawa has a flatter spectral slope and lower SMFe$^0$ content (~0.2 wt.\%) than our irradiated powder (~0.29 wt.\%), but the former has a higher $\Delta\eta$ value of ~0.25 \citep{sanchez2024population}. Considering that we already have a large amount of evidence that Kamo`oalewa is much more weathered than Itokawa \citep{zhang2025tianwen}, we remind readers to be cautious in using $\Delta\eta$ to assess the space weathering degree of S-complex asteroids. This result may be because the spectral data used to calculate \textit{PC1’} and \textit{PC2’} have removed the spectral slope information and only reflect the absorption characteristics \citep{demeo2009extension}. Therefore, we still hold the view that Kamo`oalewa is highly space-weathered and suggest that the spectral type of Kamo`oalewa be classified as Sqw type.

\section{Discussion}\label{5}
\subsection{Regolith size, thickness, and thermal inertia}\label{5.1}
Our calculated rotation period of Kamo`oalewa (27.66 min) is close to the previously reported 28.3$^{+1.8}_{-1.3}$ min \citep{sharkey2021lunar}, indicating that Kamo`oalewa is indeed a rapidly rotating object. Our global regolith critical size distribution results indicate that 93.83\% of the global surface of Kamo`oalewa is covered by grains < 2~cm in diameter, of which 67.16\% is covered by fine grains smaller than 1~cm in size. These values of critical sizes may be even smaller if the particle loss caused by grain sliding from the poles toward the equator \citep{hyodo2022formation} is also considered, consistent with another calculation work that grains smaller than 1~cm in size can be retained on Kamo`oalewa’s surface \citep{ren2024surface}. This is also consistent with the fine-sized laser-irradiated LL chondrite powder (size < 45~$\mu$m) (Fig. \ref{figA1}) rather than the slab (Fig. \ref{figA2}) showing a similar spectrum to Kamo`oalewa. The low thermal inertia (150$^{+90}_{-45}$ or 181$^{+95}_{-60}$ J m$^{-2}$ K$^{-1}$ s$^{-1/2}$) derived from the Yarkovsky effect observation also indicates that the regolith size of Kamo`oalewa is small, about 100~$\mu$m to 3~mm \citep{fenucci2025astrometry}. This is close to our laboratory measurements of irradiated LL chondrite powder, which shows a slightly lower thermal inertia (95.52 to 135.09 J m$^{-2}$ K$^{-1}$ s$^{-1/2}$) and has a slightly smaller particle size (< 45~$\mu$m). Please note that the critical size we obtained by the ``gravity-adhesion-centrifugal force'' method only represents the upper limit of regolith grains. That is, grains below this size can be retained, while the grain size obtained from the thermal inertia in \cite{fenucci2025astrometry} and our laboratory measurement represents an average size. Again, considering that our irradiated powder (< 45~$\mu$m) rather than coarse slab showed the best matching spectrum with Kamo`oalewa, these results strongly imply that the surface of Kamo`oalewa is covered by a fine regolith rather than coarse, and the grain in most surface areas is most likely between micrometers and millimeters, and will not exceed 2~cm. These also imply that Kamo`oalewa's thermal inertia is low, likely between 95.52 J m$^{-2}$ K$^{-1}$ s$^{-1/2}$ (the lower limit from our experiment) and 181$^{+95}_{-60}$ J m$^{-2}$ K$^{-1}$ s$^{-1/2}$ (the upper limit from observation of \cite{fenucci2025astrometry}).

 It is worth noting that the earlier study based on ellipsoid modeling and the ``gravity, cohesive force, centrifugal force'' balance method inferred that Kamo`oalewa’s regolith sizes are larger \citep{li2021shape}. The primary reason for the difference in our results is that \cite{li2021shape} used a theoretically derived value for the cohesive force, whereas we used a value obtained from laboratory measurements on meteorites. As we mentioned in Section \ref{3.2}, conventional models overestimate cohesive forces and fail to explain the high mobility of the asteroid surface. \cite{li2021shape} used higher values for cohesive force, resulting in an overestimation of the particle size.  It is also worth noting that \cite{liu2024surface} gave a higher thermal inertial (402.05$^{+376.29}_{-194.37}$ J m$^{-2}$ K$^{-1}$ s$^{-1/2}$) than \cite{fenucci2025astrometry} and this study. This leads them to believe that the surface particles of Kamo`oalewa are coarse, up to decimeters. \cite{liu2024surface} used a similar method to \cite{fenucci2025astrometry} (this method was developed by Fenucci et al), but the latter used their latest observational orbital data. Of particular note, two small and faster-spinning asteroids (2011 PT and 2016 GE1, with rotation periods of 10 min and 34 s) were also previously inferred to have extremely low thermal inertia or thermal conductivity, suggesting the presence of a fine-grained and thin regolith \citep{fenucci2021low,fenucci2023low}. Additionally, this study gives the first constraint on the regolith thickness of Kamo`oalewa. The thermal skin depth of 3 to 3.5~mm indicates that its regolith thickness is at least 3~mm, which is conducive to the spacecraft performing sampling operations. Current research on small and rapidly rotating asteroids is very limited, but studies on 2011 PT and 2016 GE1, together with our above work on Kamo`oalewa, suggest that such asteroids may be more likely to be covered by fine-grained and thin regolith, exhibiting low thermal inertia. In the future, Tianwen-2 probe detection of Kamo`oalewa and more survey projects are expected to test this prediction. 
\subsection{SMFe$^0$ content and spectral type}\label{5.2}
SMFe$^0$ is a key product of space weathering of silicate-rich bodies, such as the Moon and S-complex asteroids \citep{pieters2016space}, and its content is an important indicator for evaluating the degree of space weathering. Currently, a model for evaluating the degree of lunar space weathering using SMFe$^0$ content has been put into use \citep{lu2023mature}. The production rate of SMFe$^0$ on the Moon has also been studied \citep{tai2021new}. However, due to the diversities of asteroid compositions \citep{demeo2009extension,binzel2019compositional,sanchez2024population}, the diversities of space weathering effects \citep{zhang2022diverse}, and the scarcity of return missions, the rate at which SMFe$^0$ is produced by space weathering on S-complex asteroids is currently unclear. Further, the space weathering degree evaluation scheme for asteroids based on SMFe$^0$ content has not yet been established. In this study, for the first time, we predicted the SMFe$^0$ content in Kamo`oalewa's regolith, 0.29 $\pm$ 0.05 wt.\%. Considering that Kamo`oalewa and Itokawa have the same composition of LL chondrite and a similar space environment (which means a similar space weathering rate), the higher SMFe$^0$ content in Kamo`oalewa than in Itokawa implies that the former has a higher degree of space weathering than the latter.

Additionally, due to the lack of NIR data with higher spectral resolution, the spectral type of Kamo`oalewa cannot be better determined currently. However, our investigation of the virtual spectral type of laser-irradiated LL chondrite powder suggests that Sqw has high potential for Kamo`oalewa. This is consistent with Kamo`oalewa having a higher degree of space weathering than the Sq-type Itokawa. Note that, as we mentioned in Section \ref{4.6}, the Sqw is a non-standard type defined in the Bus-DeMeo taxonomy system, only indicating the presence of space weathering on Sq-type bodies. If using the Bus taxonomy system (has no ``-w'' term) or standard types in the Bus-DeMeo system, Kamo`oalewa may be classified as S-type. In the future, the data obtained by the visible-infrared spectrometer onboard the Tianwen-2 probe will bring the final answer about its spectral type.
\section{Conclusions}\label{6}
In this study, we modeled the shape, rotation period, and pole orientation of Kamo`oalewa. We also estimated the regolith critical size, regolith thickness, and surface thermal inertia of Kamo`oalewa. Furthermore, we calculated Kamo`oalewa’s SMFe$^0$ content in regolith and predicted its spectral type. We conclude that:

1. Kamo`oalewa is a sub-hundred-meter rapidly rotating asteroid, with a size of 68~m $\times$ 46~m $\times$ 39~m, a rotation period of 27.66~minutes, and a pole orientation of 134.7$^{\circ}$ longitude and $-$11.4$^{\circ}$ latitude.

2. Kamo`oalewa probably has a low surface thermal inertia, from 95 to 181$^{+95}_{-60}$ J m$^{-2}$ K$^{-1}$ s$^{-1/2}$.

3. Kamo`oalewa probably developed a fine-grained regolith, where the grain in most surface areas is between micrometers and millimeters, and will not exceed 
2 cm. 

4. The regolith depth of Kamo`oalewa is at least 3~mm.

5. Kamo`oalewa probably has a spectral type of Sqw and a high SMFe$^0$ content of 0.29 $\pm$ 0.05 wt.\% in its regolith.

\begin{acknowledgements}
We thank Yuuya Nagaashi, Haoxuan Jiang, Liangliang Yu, and Chenyang Huang for beneficial discussions on the asteroid surface state. We also thank Pei Ma for his beneficial discussion on SMFe$^0$ content estimation. Yang Li wants to thank funding support from ``From 0 to 1” Original Exploration Cultivation Project, Institute of Geochemistry, Chinese Academy of Sciences grant DHSZZ2023-3; Guizhou Provincial Foundation for Excellent Scholars Program grant GCC (2023) 088; the Youth Innovation Promotion Association of the Chinese Academy of Sciences grant 2020395; Strategic Priority Research Program of the Chinese Academy of Sciences grant XDB 41000000; and National Natural Science Foundation of China grant 42273042 and 41931077. Edward Cloutis wants to thank funding support from the Natural Sciences and Engineering Research Council of Canada grant RGPIN-2021-02995 and the Canadian Space Agency grant 22EXPOSIWI. Xiaoping Zhang wants to thank the funding support from the Science and Technology Development Fund (FDCT) of Macau, grant 0008/2024/ITP1.
\end{acknowledgements}
\bibliography{ref.bib}

\begin{thebibliography}{77}
\expandafter\ifx\csname natexlab\endcsname\relax\def\natexlab#1{#1}\fi

\bibitem[{Abe {et~al.}(2006)Abe, Takagi, Kitazato, Abe, Hiroi, Vilas, Clark, Abell, Lederer, Jarvis, {et~al.}}]{abe2006near}
Abe, M., Takagi, Y., Kitazato, K., {et~al.} 2006, Science, 312, 1334

\bibitem[{Bell {et~al.}(1988)Bell, Owensby, Hawke, \& Gaffey}]{bell198852}
Bell, J., Owensby, P., Hawke, B., \& Gaffey, M. 1988, in Abstracts of the Lunar and Planetary Science Conference, volume 19, page 57,(1988), Vol.~19

\bibitem[{Binzel {et~al.}(2019)Binzel, DeMeo, Turtelboom, Bus, Tokunaga, Burbine, Lantz, Polishook, Carry, Morbidelli, {et~al.}}]{binzel2019compositional}
Binzel, R., DeMeo, F., Turtelboom, E., {et~al.} 2019, Icarus, 324, 41

\bibitem[{Binzel {et~al.}(2001)Binzel, Rivkin, Bus, Sunshine, \& Burbine}]{binzel2001muses}
Binzel, R.~P., Rivkin, A.~S., Bus, S.~J., Sunshine, J.~M., \& Burbine, T.~H. 2001, Meteoritics \& Planetary Science, 36, 1167

\bibitem[{Binzel {et~al.}(2004)Binzel, Rivkin, Stuart, Harris, Bus, \& Burbine}]{binzel2004observed}
Binzel, R.~P., Rivkin, A.~S., Stuart, J.~S., {et~al.} 2004, Icarus, 170, 259

\bibitem[{Bro{\v{z}} {et~al.}(2024)Bro{\v{z}}, Vernazza, Marsset, DeMeo, Binzel, Vokrouhlick{\`y}, \& Nesvorn{\`y}}]{brovz2024young}
Bro{\v{z}}, M., Vernazza, P., Marsset, M., {et~al.} 2024, Nature, 634, 566

\bibitem[{Brunetto {et~al.}(2006)Brunetto, Vernazza, Marchi, Birlan, Fulchignoni, Orofino, \& Strazzulla}]{brunetto2006modeling}
Brunetto, R., Vernazza, P., Marchi, S., {et~al.} 2006, Icarus, 184, 327

\bibitem[{Bus \& Binzel(2002{\natexlab{a}})}]{bus2002phaseb}
Bus, S.~J. \& Binzel, R.~P. 2002{\natexlab{a}}, Icarus, 158, 146

\bibitem[{Bus \& Binzel(2002{\natexlab{b}})}]{bus2002phasea}
Bus, S.~J. \& Binzel, R.~P. 2002{\natexlab{b}}, Icarus, 158, 106

\bibitem[{Cahill {et~al.}(2019)Cahill, Blewett, Nguyen, Boosalis, Lawrence, \& Denevi}]{cahill2019optical}
Cahill, J.~T., Blewett, D.~T., Nguyen, N.~V., {et~al.} 2019, Icarus, 317, 229

\bibitem[{Carvano {et~al.}(2010)Carvano, Hasselmann, Lazzaro, \& Moth{\'e}-Diniz}]{carvano2010sdss}
Carvano, J., Hasselmann, P., Lazzaro, D., \& Moth{\'e}-Diniz, T. 2010, Astronomy \& Astrophysics, 510, A43

\bibitem[{Castro-Cisneros {et~al.}(2023)Castro-Cisneros, Malhotra, \& Rosengren}]{castro2023lunar}
Castro-Cisneros, J.~D., Malhotra, R., \& Rosengren, A.~J. 2023, Communications Earth \& Environment, 4, 372

\bibitem[{Chapman \& Gaffey(1979)}]{chapman1979reflectance}
Chapman, C. \& Gaffey, M. 1979, Asteroids, 655

\bibitem[{Cloutis {et~al.}(2014)Cloutis, Binzel, \& Gaffey}]{cloutis2014establishing}
Cloutis, E.~A., Binzel, R.~P., \& Gaffey, M.~J. 2014, Elements, 10, 25

\bibitem[{De~la Fuente~Marcos \& De~la Fuente~Marcos(2016)}]{de2016asteroid}
De~la Fuente~Marcos, C. \& De~la Fuente~Marcos, R. 2016, Monthly Notices of the Royal Astronomical Society, 462, 3441

\bibitem[{de~Leon {et~al.}(2018)de~Leon, Pinilla-Alonso, Campins, Licandro, Morate, Lorenzi, De~Pr{\'a}, \& Rizos}]{de2018primitive}
de~Leon, J., Pinilla-Alonso, N., Campins, H., {et~al.} 2018, in AAS/Division for Planetary Sciences Meeting Abstracts\# 50, Vol.~50, 310--05

\bibitem[{DeMeo {et~al.}(2009)DeMeo, Binzel, Slivan, \& Bus}]{demeo2009extension}
DeMeo, F.~E., Binzel, R.~P., Slivan, S.~M., \& Bus, S.~J. 2009, Icarus, 202, 160

\bibitem[{DeMeo {et~al.}(2022)DeMeo, Burt, Marsset, Polishook, Burbine, Carry, Binzel, Vernazza, Reddy, Tang, {et~al.}}]{demeo2022connecting}
DeMeo, F.~E., Burt, B.~J., Marsset, M., {et~al.} 2022, Icarus, 380, 114971

\bibitem[{DeMeo \& Carry(2014)}]{demeo2014solar}
DeMeo, F.~E. \& Carry, B. 2014, Nature, 505, 629

\bibitem[{DeMeo {et~al.}(2023)DeMeo, Marsset, Polishook, Burt, Binzel, Hasegawa, Granvik, Moskovitz, Earle, Bus, {et~al.}}]{demeo2023isolating}
DeMeo, F.~E., Marsset, M., Polishook, D., {et~al.} 2023, Icarus, 389, 115264

\bibitem[{Egan \& Hilgeman(1977)}]{egan1977rings}
Egan, W. \& Hilgeman, T. 1977, Icarus, 30, 413

\bibitem[{Fenucci \& Novakovi{\'c}(2021)}]{fenucci2021role}
Fenucci, M. \& Novakovi{\'c}, B. 2021, The Astronomical Journal, 162, 227

\bibitem[{Fenucci {et~al.}(2026)Fenucci, Novakovi{\'c}, Granvik, \& Zhang}]{fenucci2026origin}
Fenucci, M., Novakovi{\'c}, B., Granvik, M., \& Zhang, P. 2026, arXiv preprint arXiv:2601.08923

\bibitem[{Fenucci {et~al.}(2023)Fenucci, Novakovi{\'c}, \& Mar{\v{c}}eta}]{fenucci2023low}
Fenucci, M., Novakovi{\'c}, B., \& Mar{\v{c}}eta, D. 2023, Astronomy \& Astrophysics, 675, A134

\bibitem[{Fenucci {et~al.}(2021)Fenucci, Novakovi{\'c}, Vokrouhlick{\`y}, \& Weryk}]{fenucci2021low}
Fenucci, M., Novakovi{\'c}, B., Vokrouhlick{\`y}, D., \& Weryk, R.~J. 2021, Astronomy \& Astrophysics, 647, A61

\bibitem[{Fenucci {et~al.}(2025)Fenucci, Novakovi{\'c}, Zhang, Carbognani, Micheli, Faggioli, Gianotto, Oca{\~n}a, F{\"o}hring, Cano, {et~al.}}]{fenucci2025astrometry}
Fenucci, M., Novakovi{\'c}, B., Zhang, P., {et~al.} 2025, Astronomy \& Astrophysics, 695, A196

\bibitem[{Galluccio {et~al.}(2023)Galluccio, Delbo, De~Angeli, Pauwels, Tanga, Mignard, Cellino, Brown, Muinonen, Penttil{\"a}, {et~al.}}]{galluccio2023gaia}
Galluccio, L., Delbo, M., De~Angeli, F., {et~al.} 2023, Astronomy \& Astrophysics, 674, A35

\bibitem[{Granvik {et~al.}(2018)Granvik, Morbidelli, Jedicke, Bolin, Bottke, Beshore, Vokrouhlick{\`y}, Nesvorn{\`y}, \& Michel}]{granvik2018debiased}
Granvik, M., Morbidelli, A., Jedicke, R., {et~al.} 2018, Icarus, 312, 181

\bibitem[{Greenwood {et~al.}(2023)Greenwood, Franchi, Findlay, Malley, Ito, Yamaguchi, Kimura, Tomioka, Uesugi, Imae, {et~al.}}]{greenwood2023oxygen}
Greenwood, R.~C., Franchi, I.~A., Findlay, R., {et~al.} 2023, Nature Astronomy, 7, 29

\bibitem[{Hapke(1981)}]{hapke1981bidirectional}
Hapke, B. 1981, Journal of Geophysical Research: Solid Earth, 86, 3039

\bibitem[{Hapke(2001)}]{hapke2001space}
Hapke, B. 2001, Journal of Geophysical Research: Planets, 106, 10039

\bibitem[{Hapke(2012)}]{hapke2012theory}
Hapke, B. 2012, Theory of reflectance and emittance spectroscopy (Cambridge university press)

\bibitem[{Hiroi {et~al.}(2006)Hiroi, Abe, Kitazato, Abe, Clark, Sasaki, Ishiguro, \& Barnouin-Jha}]{hiroi2006developing}
Hiroi, T., Abe, M., Kitazato, K., {et~al.} 2006, Nature, 443, 56

\bibitem[{Hyodo \& Sugiura(2022)}]{hyodo2022formation}
Hyodo, R. \& Sugiura, K. 2022, The Astrophysical Journal Letters, 937, L36

\bibitem[{Jiao {et~al.}(2024)Jiao, Cheng, Huang, Asphaug, Gladman, Malhotra, Michel, Yu, \& Baoyin}]{jiao2024asteroid}
Jiao, Y., Cheng, B., Huang, Y., {et~al.} 2024, Nature Astronomy, 1

\bibitem[{Jin \& Ishiguro(2022)}]{jin2022estimation}
Jin, S. \& Ishiguro, M. 2022, Astronomy \& Astrophysics, 667, A93

\bibitem[{Keller {et~al.}(2025)Keller, Thompson, Seifert, Melendez, Thomas-Keprta, Le, Snead, Welten, Nishiizumi, Caffee, {et~al.}}]{keller2025space}
Keller, L.~P., Thompson, M.~S., Seifert, L.~B., {et~al.} 2025, Nature Geoscience, 1

\bibitem[{Lauretta {et~al.}(2017)Lauretta, Balram-Knutson, Beshore, Boynton, Drouet~d’Aubigny, DellaGiustina, Enos, Golish, Hergenrother, Howell, {et~al.}}]{lauretta2017osiris}
Lauretta, D., Balram-Knutson, S., Beshore, E., {et~al.} 2017, Space Science Reviews, 212, 925

\bibitem[{Lawrence \& Lucey(2007)}]{lawrence2007radiative}
Lawrence, S.~J. \& Lucey, P.~G. 2007, Journal of Geophysical Research: Planets, 112

\bibitem[{Levison {et~al.}(2021)Levison, Olkin, Noll, Marchi, Bell~III, Bierhaus, Binzel, Bottke, Britt, Brown, {et~al.}}]{levison2021lucy}
Levison, H.~F., Olkin, C.~B., Noll, K.~S., {et~al.} 2021, The Planetary Science Journal, 2, 171

\bibitem[{Li \& Scheeres(2021)}]{li2021shape}
Li, X. \& Scheeres, D.~J. 2021, Icarus, 357, 114249

\bibitem[{Liu {et~al.}(2024)Liu, Chen, Yan, Yu, Fenucci, Ye, Zhong, Qiu, \& Barriot}]{liu2024surface}
Liu, L., Chen, Q., Yan, J., {et~al.} 2024, Solar System Research, 58, 469

\bibitem[{Lord {et~al.}(2017)Lord, Tilley, Oh, Goebel, Polanskey, Snyder, Carr, Collins, Lantoine, Landau, {et~al.}}]{lord2017psyche}
Lord, P., Tilley, S., Oh, D.~Y., {et~al.} 2017, in 2017 IEEE Aerospace Conference, IEEE, 1--11

\bibitem[{Lu {et~al.}(2023)Lu, Chen, Ling, Liu, Fu, Qiao, Zhang, Cao, Liu, He, {et~al.}}]{lu2023mature}
Lu, X., Chen, J., Ling, Z., {et~al.} 2023, Nature Astronomy, 7, 142

\bibitem[{Lu {et~al.}(2014)Lu, Zhao, \& You}]{lu2014cellinoid}
Lu, X., Zhao, H., \& You, Z. 2014, Earth, Moon, and Planets, 112, 73

\bibitem[{Lucey(1998)}]{lucey1998model}
Lucey, P.~G. 1998, Journal of Geophysical Research: Planets, 103, 1703

\bibitem[{Lucey \& Riner(2011)}]{lucey2011optical}
Lucey, P.~G. \& Riner, M.~A. 2011, Icarus, 212, 451

\bibitem[{Macke(2010)}]{macke2010survey}
Macke, R. 2010, Orlando, FL: University of Central Florida

\bibitem[{Marchi {et~al.}(2005)Marchi, Brunetto, Magrin, Lazzarin, \& Gandolfi}]{marchi2005space}
Marchi, S., Brunetto, R., Magrin, S., Lazzarin, M., \& Gandolfi, D. 2005, Astronomy \& Astrophysics, 443, 769

\bibitem[{Marchi {et~al.}(2006)Marchi, Magrin, Nesvorn{\`y}, Paolicchi, \& Lazzarin}]{marchi2006spectral}
Marchi, S., Magrin, S., Nesvorn{\`y}, D., Paolicchi, P., \& Lazzarin, M. 2006, Monthly Notices of the Royal Astronomical Society: Letters, 368, L39

\bibitem[{Marty {et~al.}(2025)Marty, Zimmermann, F{\"u}ri, Bekaert, Barnes, Nguyen, Connolly, \& Lauretta}]{marty2025noble}
Marty, B., Zimmermann, L., F{\"u}ri, E., {et~al.} 2025, Meteoritics \& Planetary Science

\bibitem[{Nagaashi \& Nakamura(2023)}]{nagaashi2023high}
Nagaashi, Y. \& Nakamura, A.~M. 2023, Science Advances, 9, eadd3530

\bibitem[{Nagao {et~al.}(2011)Nagao, Okazaki, Nakamura, Miura, Osawa, Bajo, Matsuda, Ebihara, Ireland, Kitajima, {et~al.}}]{nagao2011irradiation}
Nagao, K., Okazaki, R., Nakamura, T., {et~al.} 2011, Science, 333, 1128

\bibitem[{Nakamura {et~al.}(2011)Nakamura, Noguchi, Tanaka, Zolensky, Kimura, Tsuchiyama, Nakato, Ogami, Ishida, Uesugi, {et~al.}}]{nakamura2011itokawa}
Nakamura, T., Noguchi, T., Tanaka, M., {et~al.} 2011, Science, 333, 1113

\bibitem[{Nesvorn{\`y} {et~al.}(2005)Nesvorn{\`y}, Jedicke, Whiteley, \& Ivezi{\'c}}]{nesvorny2005evidence}
Nesvorn{\`y}, D., Jedicke, R., Whiteley, R.~J., \& Ivezi{\'c}, {\v{Z}}. 2005, Icarus, 173, 132

\bibitem[{Noguchi {et~al.}(2023)Noguchi, Matsumoto, Miyake, Igami, Haruta, Saito, Hata, Seto, Miyahara, Tomioka, {et~al.}}]{noguchi2023dehydrated}
Noguchi, T., Matsumoto, T., Miyake, A., {et~al.} 2023, Nature Astronomy, 7, 170

\bibitem[{Okazaki {et~al.}(2022)Okazaki, Miura, Takano, Sawada, Sakamoto, Yada, Yamada, Kawagucci, Matsui, Hashizume, {et~al.}}]{okazaki2022first}
Okazaki, R., Miura, Y.~N., Takano, Y., {et~al.} 2022, Science advances, 8, eabo7239

\bibitem[{Pieters \& Noble(2016)}]{pieters2016space}
Pieters, C.~M. \& Noble, S.~K. 2016, Journal of Geophysical Research: Planets, 121, 1865

\bibitem[{Prockter {et~al.}(2002)Prockter, Murchie, Cheng, Krimigis, Farquhar, Santo, \& Trombka}]{prockter2002near}
Prockter, L., Murchie, S., Cheng, A., {et~al.} 2002, Acta Astronautica, 51, 491

\bibitem[{Ren {et~al.}(2024)Ren, Wu, Hesse, Li, Liu, \& Wang}]{ren2024surface}
Ren, J., Wu, B., Hesse, M.~A., {et~al.} 2024, Astronomy \& Astrophysics, 692, A62

\bibitem[{Russell \& Raymond(2011)}]{russell2011dawn}
Russell, C.~T. \& Raymond, C.~A. 2011, Space Science Reviews, 163, 3

\bibitem[{Sanchez {et~al.}(2024)Sanchez, Reddy, Thirouin, Bottke, Kareta, De~Florio, Sharkey, Battle, Cantillo, \& Pearson}]{sanchez2024population}
Sanchez, J.~A., Reddy, V., Thirouin, A., {et~al.} 2024, The Planetary Science Journal, 5, 131

\bibitem[{Sasaki {et~al.}(2001)Sasaki, Nakamura, Hamabe, Kurahashi, \& Hiroi}]{sasaki2001production}
Sasaki, S., Nakamura, K., Hamabe, Y., Kurahashi, E., \& Hiroi, T. 2001, Nature, 410, 555

\bibitem[{Scheeres {et~al.}(2010)Scheeres, Hartzell, S{\'a}nchez, \& Swift}]{scheeres2010scaling}
Scheeres, D.~J., Hartzell, C.~M., S{\'a}nchez, P., \& Swift, M. 2010, Icarus, 210, 968

\bibitem[{Sei-ichiro {et~al.}(2017)Sei-ichiro, Yuichi, Makoto, Tanaka, Takanao, \& Satoru}]{sei2017hayabusa2}
Sei-ichiro, W., Yuichi, T., Makoto, Y., {et~al.} 2017, Space Science Reviews, 208, 3

\bibitem[{Sharkey {et~al.}(2021)Sharkey, Reddy, Malhotra, Thirouin, Kuhn, Conrad, Rothberg, Sanchez, Thompson, \& Veillet}]{sharkey2021lunar}
Sharkey, B.~N., Reddy, V., Malhotra, R., {et~al.} 2021, Communications Earth \& Environment, 2, 231

\bibitem[{Tai~Udovicic {et~al.}(2021)Tai~Udovicic, Costello, Ghent, \& Edwards}]{tai2021new}
Tai~Udovicic, C., Costello, E., Ghent, R., \& Edwards, C. 2021, Geophysical Research Letters, 48, e2020GL092198

\bibitem[{Trang {et~al.}(2013)Trang, Lucey, Gillis-Davis, Cahill, Klima, \& Isaacson}]{trang2013near}
Trang, D., Lucey, P.~G., Gillis-Davis, J.~J., {et~al.} 2013, Journal of Geophysical Research: Planets, 118, 708

\bibitem[{Vernazza {et~al.}(2009)Vernazza, Binzel, Rossi, Fulchignoni, \& Birlan}]{vernazza2009solar}
Vernazza, P., Binzel, R., Rossi, A., Fulchignoni, M., \& Birlan, M. 2009, Nature, 458, 993

\bibitem[{Wang {et~al.}(2026)Wang, Hu, Ji, \& Ying}]{wang2026dynamical}
Wang, Y., Hu, S., Ji, J., \& Ying, J.-j. 2026, Research in Astronomy and Astrophysics

\bibitem[{Warell \& Davidsson(2010)}]{warell2010hapke}
Warell, J. \& Davidsson, B. 2010, Icarus, 209, 164

\bibitem[{Yoshikawa {et~al.}(2021)Yoshikawa, Kawaguchi, Fujiwara, \& Tsuchiyama}]{yoshikawa2021hayabusa}
Yoshikawa, M., Kawaguchi, J., Fujiwara, A., \& Tsuchiyama, A. 2021, in Sample return missions (Elsevier), 123--146

\bibitem[{Yurimoto {et~al.}(2011)Yurimoto, Abe, Abe, Ebihara, Fujimura, Hashiguchi, Hashizume, Ireland, Itoh, Katayama, {et~al.}}]{yurimoto2011oxygen}
Yurimoto, H., Abe, K.-i., Abe, M., {et~al.} 2011, Science, 333, 1116

\bibitem[{Zellner {et~al.}(1985)Zellner, Tholen, \& Tedesco}]{zellner1985eight}
Zellner, B., Tholen, D.~J., \& Tedesco, E. 1985, Icarus, 61, 355

\bibitem[{Zhang {et~al.}(2022)Zhang, Tai, Li, Zhang, Lantz, Hiroi, Matsuoka, Li, Lin, Wen, {et~al.}}]{zhang2022diverse}
Zhang, P., Tai, K., Li, Y., {et~al.} 2022, Astronomy \& Astrophysics, 659, A78

\bibitem[{Zhang {et~al.}(2025{\natexlab{a}})Zhang, Zhang, Wei, Granvik, Yan, Wang, Zhang, Pang, Jiang, Vernazza, {et~al.}}]{zhang2025tianwen}
Zhang, P., Zhang, G., Wei, Z., {et~al.} 2025{\natexlab{a}}, Nature Communications, Under Review

\bibitem[{Zhang {et~al.}(2025{\natexlab{b}})Zhang, Guo, Zheng, \& Wang}]{zhang2025dcappso}
Zhang, Y.-X., Guo, W.-X., Zheng, H., \& Wang, W.-L. 2025{\natexlab{b}}, Astronomy and Computing, 51, 100925

\end{thebibliography}
\begin{appendix}

\section{Identification of the componsition of Kamo`oalewa} \label{appendixA}
\begin{figure*}[h!]
    \centering
   \includegraphics[width=18 cm]{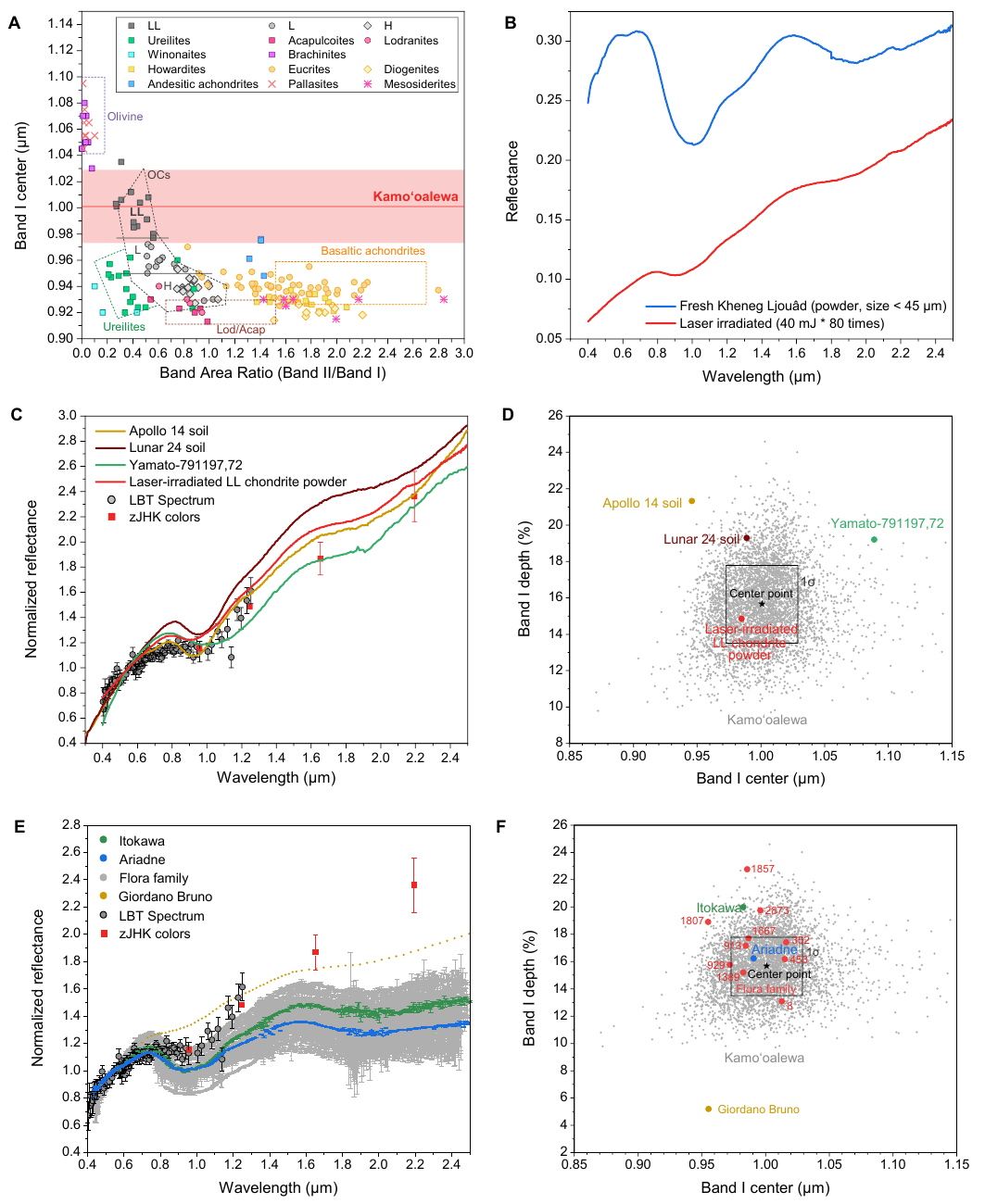}
     \caption{Spectral evidence for identifying the composition of Kamo`oalewa. (A) Band parameters of Kamo`oalewa and meteorites. (B) Reflectance spectra of fresh and laser-irradiated LL chondrite powder. (C) 0.55~$\mu$m-normalized spectra of Kamo`oalewa, three lunar samples, and laser-irradiated LL chondrite powder. (D) Band parameters of Kamo`oalewa, three lunar samples, and laser-irradiated LL chondrite powder. (E) Reflectance spectra of Kamo`oalewa, Itokawa, Flora family, and the lunar Giordano Bruno crater. (F) Band parameters of Kamo`oalewa, Itokawa, the Flora family, and the lunar Giordano Bruno crater.}
     \label{figA1}
     \end{figure*}

\begin{figure*}[h!]
    \centering
   \includegraphics[width=9 cm]{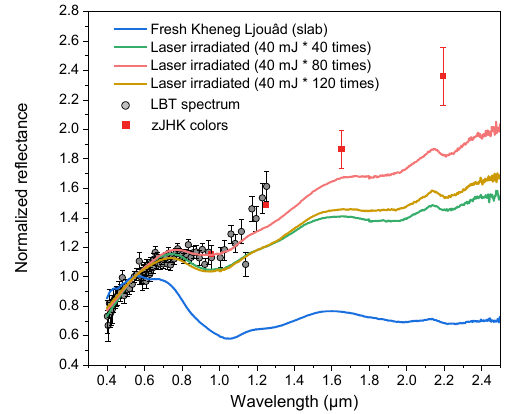}
     \caption{0.55~$\mu$m-normalized spectra of Kamo`oalewa and fresh and laser-irradiated LL chondrite slab.}
     \label{figA2}
     \end{figure*}

In our previous study \citep{zhang2025tianwen}, we conducted a combined spectroscopic and orbital-dynamical analysis of Kamo`oalewa. Fig. \ref{figA1} shows our part spectral results. After artificially generating 10000 spectra by the Monte Carlo method for the low signal-to-noise ratio spectrum of Kamo`oalewa and removing the continuum, we obtained a statistical Band I center of 1.001±0.028~$\mu$m (error is 1$\sigma$), falling into the range of LL chondrites (Fig. \ref{figA1}A). Due to the Band I center is diagnostic of composition, 1.001±0.028~$\mu$m indicates that Kamo`oalewa is similar to LL chondrites in composition. Since Kamo`oalewa shows an extremely red spectral slope \citep{sharkey2021lunar}, we speculate that it may have undergone intense space weathering. To test this scenario, we conducted irradiation experiments on an LL5/6 ordinary chondrite (Kheneg Ljouâd) using a nanosecond pulse laser, aiming to simulate space weathering effects. For a powder sample with a size < 45~$\mu$m, we irradiated it at different energies. As shown in Fig. \ref{figA1}B, when the energy was set at 40 mJ $\times$ 80 times, we obtain a spectral curve with a steep slope and a reflectance at 0.55~$\mu$m of 0.085. Similar to several earlier reports of lunar materials, this spectral curve also matches Kamo`oalewa (Fig. \ref{figA1}C). We also irradiated a slab sample of Kheneg Ljouâd; however, even if space weathering is saturated (the spectral slope is no longer increasing), it cannot produce a spectrum that matches Kamo`oalewa (Fig. \ref{figA2}). Therefore, Fig. \ref{figA1}A-D and Fig. \ref{figA2} suggest that, if Kamo`oalewa has a composition of LL chondrite and has undergone space weathering, then the surface regolith should be fine rather than coarse. 

The previous two studies \citep{sharkey2021lunar,jiao2024asteroid} found that only three lunar materials (Apollo 14 soil, Lunar 24 soil, and lunar meteorite Yamato-791197,72) exhibit spectra similar to Kamo`oalewa (\ref{figA1}C). These three samples were therefore used as spectroscopic evidence to support the lunar origin of Kamo`oalewa. However, neither of these studies performed detailed spectral parameter calculations and analyses, nor did they consider the possibility of an LL chondrite. We found that Apollo 14 soil and Yamato-791197,72 do not fall into the 1$\sigma$ range of Kamo`oalewa's Band I center (Fig. \ref{figA1}D). Although Lunar 24 soil has a Band I center similar to Kamo`oalewa (Fig. \ref{figA1}D), it shows a slightly redder spectrum than Kamo`oalewa. If Kamo`oalewa is similar to Lunar 24 soil, its bluer spectrum indicates a lower degree of space weathering, so it should show a deeper Band I depth than Lunar 24 soil. However, as shown in Fig. \ref{figA1}D, this is not the case. These three lunar materials, therefore, do not appear to be good analogues of Kamo`oalewa. Our laser-irradiated LL chondrite powder shows a Band I center matched to Kamo`oalewa (Fig. \ref{figA1}D). It also has a slightly redder spectral slope than Kamo`oalewa (Fig. \ref{figA1}C). If Kamo`oalewa has an LL chondrite composition, its bluer spectral slope means a lower degree of space weathering, and it should exhibit a deeper Band I depth than our laser-irradiated LL chondrite powder. As shown in Fig. \ref{figA1}D, this is indeed the case. These suggest that the LL chondrite is a more plausible composition of Kamo`oalewa than these three lunar materials. As mentioned in section \ref{2}, our orbital dynamics model suggests that Kamo`oalewa likely originated from the Flora family close to the inner main belt when assuming it is an LL chondrite compositional object, while \cite{jiao2024asteroid} previously proposed that if Kamo`oalewa originated from the Moon, the Bruno crater is the most likely source region. To further trace the source region and composition of Kamo`oalewa, we also compare the spectra of Kamo`oalewa, Itokawa, Bruno crater, and FLora family bodies. As shown in Fig. \ref{figA1}E, the Bruno crater shows weak Band I absorption and a red spectral slope, indicating that it has experienced a certain degree of space weathering. If Kamo`oalewa originated from the Bruno crater, the deeper Band I absorption of the former than the latter means a higher degree of space weathering on Kamo`oalewa, then the spectral slope of Kamo`oalewa should be less red than the  Bruno crater, but that is not the case (Fig. \ref{figA1}F). Additionally, the Band I center of Bruno crater does not fall into the 1$\sigma$ range of Kamo`oalewa. The Bruno crater, therefore, is implausibly the source region of Kamo`oalewa. If, as revealed by \cite{jiao2024asteroid}, dynamical simulations support that the Bruno crater is the most likely source region of Kamo`oalewa, then Kamo`oalewa shouldn't originate from the Moon. Itokawa and the 25 Flora family bodies show a bluer spectral slope than Kamo`oalewa (Fig. \ref{figA1}E). Of which, Itokawa and 7 bodies in the Flora family have a Band I center within 1$\sigma$ range of Kamo`oalewa and show deeper Band I depths than Kamo`oalewa (Fig. \ref{figA1}F). These suggest that Itokawa and the 7 Flora family bodies likely have the same composition as Kamo`oalewa, but exhibit a lower degree of space weathering. The consistency of our spectroscopic and dynamical analyses therefore supports that Kamo`oalewa is composed of LL chondrite and has a highly space-weathered surface. Two recent quantitative estimates \citep{fenucci2026origin,wang2026dynamical} have suggested that Kamo`oalewa-like objects can also originate in the main belt, without relying on the lunar origin hypothesis. Meanwhile, the numerical simulation indicates that the expected number of Kamo`oalewa-like objects originating from the main belt is more than one order of magnitude greater than that originating from the Bruno crater \citep{fenucci2026origin}. Therefore, this study was conducted under the premise that Kamo`oalewa has a highly space-weathered surface with an LL chondrite composition. Please see more information in \cite{zhang2025tianwen}.

\section{Measurement of mineral types and abundance of LL5/6 chondrite Kheneg Ljouâd} \label{appendixB}
\begin{figure*}[h!]
    \centering
   \includegraphics[width=16 cm]{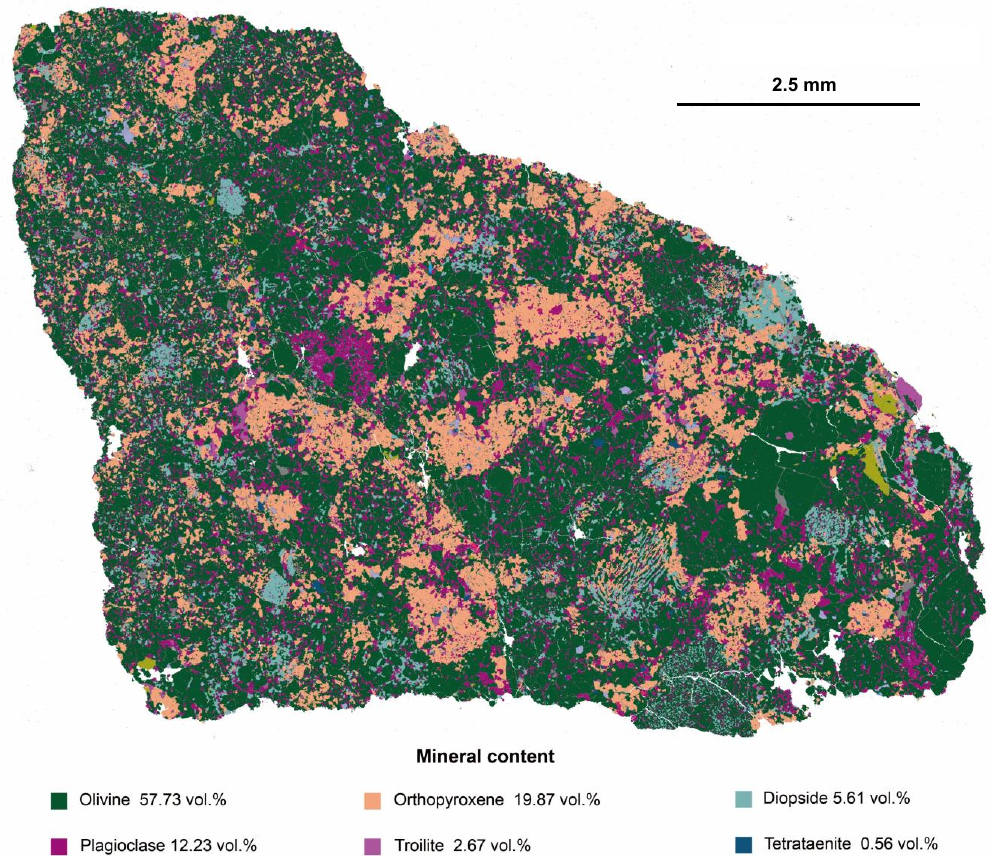}
     \caption{Mineral distribution and abundance of the LL5/6 chondrite Kheneg Ljouâd.}
     \label{figB1}
     \end{figure*}

The mineral types and abundance of LL5/6 chondrite Kheneg Ljouâd were obtained by the TESCAN Integrated Mineral Analyzer (TIMA), which is located at the Nanjing Hongchuang Geological Exploration Technology Service Co., Ltd. This equipment is equipped with a MIRA-3 scanning electron microscope and four X-ray energy dispersive spectroscopy detectors to obtain the element information. As a result, this meteorite has 57.73 vol.\% olivine, 19.87 vol.\% orthopyroxene, 5.61 vol.\% diopside, 12.23 vol.\% plagioclase, 2.62 vol.\% troilite, 0.05 vol.\% nickel-bearing troilite, 0.56 vol.\% tetrataenite, 0.76 vol.\% chromite, 0.03 vol.\% ilmenite, 0.12 vol.\% orthoclase, and 0.38 vol.\% apatite, as shown in Fig. \ref{figB1}.

The geochemistry measurements from the Meteoritical Bulletin Web\footnote{https://www.lpi.usra.edu/meteor/metbull.php} show that the mineral composition of Kheneg Ljouâd is: olivine Fa$_{31.0\pm0.2}$, low Ca pyroxene Fs$_{25.0\pm0.4}$Wo$_{2.1\pm0.2}$, high Ca pyroxene Fs$_{10.7}$Wo$_{43.1}$ and Fs$_{11.0}$Wo$_{43.0}$, Feldspar Ab$_{84.4\pm2.2}$An$_{10.6\pm0.3}$Or$_{5.0\pm2.3}$, and tetrataenite (at\%) Fe$_{42.9\pm0.2}$Co$_{2.1\pm0.1}$Ni$_{54.9\pm0.2}$.

\section{Estimation of Itokawa’s SMFe$^0$ content} \label{appendixC}
\begin{figure*}[h!]
    \centering
   \includegraphics[width=9 cm]{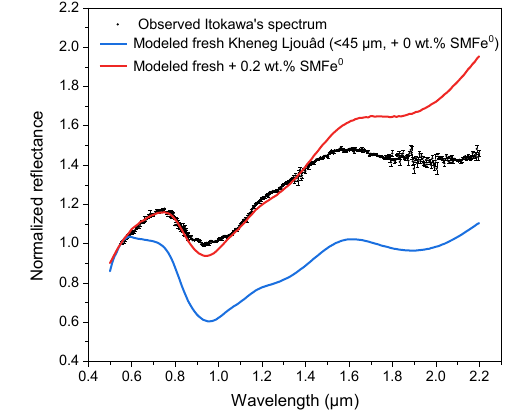}
     \caption{Estimation of Itokawa’s SMFe$^0$ content. The observed Itokawa’s spectrum best corresponds to 0.2 wt.\% SMFe$^0$ content. Itokawa’s observed spectrum is cited from \cite{binzel2001muses}. Spectra are normalized at 0.55~$\mu$m.}
     \label{figC1}
     \end{figure*}

To test the feasibility of our model, we estimated the SMFe$^0$ content in the regolith of Itokawa before calculating for Kamo`oalewa. As described in Section \ref{3.5}, we used the same program as Kamo`oalewa to calculate Itokawa’s SMFe$^0$ content, except that the mineral proportions were different. As shown in Fig. \ref{figC1}, when the SMFe$^0$ content was added to 0.2 wt.\%, the corresponding spectral curve best matched that of Itokawa. This is consistent with previous estimations \citep{binzel2001muses,hiroi2006developing}, indicating that this model can be well applied to Kamo`oalewa.

\end{appendix}
\end{document}